\documentclass[twocolumn, astrosymb, doublespacing, twocolappendix]{aastex631} 
\usepackage{amsmath}
\usepackage{mathtools}
\usepackage{appendix}
\usepackage{dsfont}
\usepackage{listings}
\usepackage{placeins}
\usepackage{booktabs}

\usepackage{hyperref}

\shorttitle{Rotation State of `Oumuamua}
\shortauthors{Taylor et al.}

\graphicspath{{./}{figures/}}

\begin{document}
\title{Fitting the Light Curve of 1I/`Oumuamua with a Nonprincipal Axis Rotational Model and Outgassing Torques}

\correspondingauthor{Aster Taylor}
\email{astertaylor@uchicago.edu}
\author[0000-0002-0140-4475]{Aster G. Taylor}
\altaffiliation{{Supported by the Fannie and John Hertz Foundation}}
\affiliation{Dept. of Astronomy and Astrophysics, University of Chicago, 5640 S Ellis Ave, Chicago, IL 60637}
\affiliation{{Dept. of Astronomy, University of Michigan, Ann Arbor, MI 48109}}

\author[0000-0002-0726-6480]{Darryl Z. Seligman}
\affiliation{Dept. of Astronomy and Carl Sagan Institute, Cornell University, 122 Sciences Drive, Ithaca, NY, 14853, USA}

\author[0000-0001-6952-9349]{Olivier R. Hainaut}
\affiliation{European Southern Observatory, Karl-Schwarzschild-Strasse 2, Garching bei München, D-85748, Germany}

\author[0000-0002-2058-5670]{Karen J. Meech}
\affil{Institute for Astronomy, University of Hawaii, 2680 Woodlawn Drive, Honolulu, HI 96822, USA}

\begin{abstract}

In this paper, we investigate the nonprincipal axis (NPA) rotational state  of  1I/`Oumuamua --- the first interstellar object discovered traversing the inner Solar System --- from its photometric light curve. Building upon  \citet{Mashchenko2019}, we develop a model which incorporates NPA rotation and {Sun-induced, time-varying} outgassing torques to generate synthetic light curves of the object. The model neglects tidal forces, which are negligible compared to outgassing torques over the distances that `Oumuamua was observed. We implement an optimization scheme that incorporates the NPA rotation model to calculate the initial rotation state of the object. We find that an NPA rotation state with an average period of $\langle P \rangle\simeq7.34$ hr best reproduces the photometric data. The discrepancy between this period and previous estimates is due to continuous period modulation induced by  outgassing torques in the rotational model, {as well as different periods being used}. The best fit to the October 2017 data does not reproduce the November 2017 data (although the later measurements are too sparse to fit). The light curve is consistent with no secular evolution of the angular momentum, somewhat in tension with the empirical correlations between nuclear spin-up and cometary outgassing. The complex rotation of `Oumuamua may be {the result of primordial rotation about the smallest principal axis} if (i) the object experienced hypervolatile outgassing and (ii) our idealized outgassing model is accurate. 

\end{abstract}

\keywords{Interstellar Objects (52); Comets (280); Light curves (918); Small Solar System bodies (1469)}

\section{Introduction}

The first interstellar object, 1I/`Oumuamua, was discovered postperihelion and reported by the Minor Planet Center in 2017 \citep{mpec2017}. Ground- and space-based observations were acquired, which produced a high-quality light curve spanning 29.3 d and 0.13 au. The 818 photometric observations were reported by \citet{meech2017}, \citet{bolin2017}, \citet{bannister2017}, \citet{drahus2017}, \citet{fraser2018}, \citet{jewitt2017}, \citet{knight2017}, and \citet{belton2018}. These observations were collected in \citet{belton2018}. 

Analysis of the light curve revealed a periodicity of $P\simeq4.3$ hr \citep{belton2018}, which was interpreted as the half-period of a rotating elongated body. The variations in the absolute magnitude indicated that the object was experiencing complex, NPA rotation \citep{drahus2017,meech2017,fraser2018}. A wide variety of aspect ratios were proposed to explain `Oumuamua's light curve variations, from $>$3:1 \citep{bolin2017,knight2017}, $>$5:1 \citep{bannister2017,fraser2018,jewitt2017}, and up to 10:1 \citep{meech2017}. 

Deep imaging revealed a notable lack of cometary activity which led to the classification of `Oumuamua as an asteroidal body and estimated upper limits on the dust production (upper limits range from $\sim2\times 10^{-4}$ kg s$^{-1}$ \citep{jewitt2017} to $1.7\times10^{-3}$ kg s$^{-1}$ \citep{meech2017}). There was a significant nondetection of the object with the \textit{Spitzer Space Telescope} outbound at 2 au which placed upper limits on the production of CO and CO$_2$ \citep{trilling2018}. 

Using a complex rotation model and full light curve modeling, \citet{Mashchenko2019} showed that a near-symmetric oblate ellipsoid with dimensions of 115:111:19 {m}$\sim$6:6:1 provided a best fit geometry for the light curve data, assuming a geometric albedo of $p=0.1$. A prolate ellipsoid with dimensions of 342:42:42 m is also allowable, but is disfavored as the torques required to replicate the motion are highly tuned. Notably, \citet{Mashchenko2019} showed that a continuous torque (and one that is constant in the comoving frame) was necessary for `Oumuamua to match the observed light curve. Notably, this torque would necessarily induce a spin-up in the rotation state of the body \citep{rafikov2018a,rafikov2018b}. 

Meanwhile, astrometric positional data revealed that the trajectory was inconsistent with pure Keplerian motion \citep{micheli2018}. The addition of a radial nongravitational acceleration of the form $a_{\rm ng}=4.92\times 10^{-4}\, (r/1\, {\rm au})^{-2}\, \boldsymbol{\hat{r}}$ cm s$^{-2}$ provides a $30\sigma$ improved fit to the trajectory.\footnote{An acceleration with exponent anywhere between $-1$ and $-2$ is equally as good of a fit and indistinguishable.} \citet{micheli2018} proposed a comet-like outgassing as an explanation for this acceleration, ruling out radiation pressure, the Yarkovsky effect, magnetic forces, a binary object, a photocenter offset, an impulsive velocity change, and friction-like effects aligned with the velocity vector. 

The stringent limits on the presence of micron-scale dust is challenging to rectify with the hypothesis that outgassing provided the nongravitational acceleration.\footnote{Although the original \textit{Spitzer} estimates had a computational error \citep{seligman2021}, even these revised CO limits are prohibitive to the $1/r^2$ fit.} Many authors have proposed alternative mechanisms to explain this behavior, including various possible ices as well as more exotic scenarios. 

While H$_2$O ice is the most common volatile in solar system comets \citep{Rickman2010,Ahearn2012,Ootsubo2012,Cochran2015,Biver2016,Bockelee2017} and its presence is not in tension with either the \textit{Spitzer} nondetection or the lack of OH detected by \citet{Park2018}, its enthalpy of sublimation is relatively high (51 kJ mol$^{-1}$). For water sublimation to provide the nongravitational acceleration, there would have to have been substantially more energy input than `Oumuamua received from solar radiation \citep{sekanina2019a}. 

\citet{SL2020} argued that only hypervolatiles could serve as the accelerant for `Oumuamua. They found that only molecular hydrogen (H$_2$), neon, molecular nitrogen (N$_2$), and argon were allowable accelerants for an oblate spheroid, although CO was also shown to be energetically feasible. Those authors also investigated the feasibility of hydrogen ice as the bulk constituent --- originally hypothesized by \citet{fuglistaler2018solid} --- as it requires the lowest active surface fraction. In this hypothesis, `Oumuamua would have formed in a failed prestellar core in a Giant Molecular Cloud. This model naturally explains many oddities of `Oumuamua, including the extreme shape (via continuous H$_2$ ablation), the low excess velocity speed, and young age \citep{mamajek2017,Gaidos2017a, Feng2018,Fernandes2018,hallatt2020,Hsieh2021}. However, the formation of macroscopic bodies composed of solid hydrogen is theoretically difficult due to the frigid temperatures required for formation and its rapid evaporation in the interstellar medium \citep{hoang2020,phan2021,levine2021,LL2021}. 

\citet{jackson2021} instead suggested that `Oumuamua was composed of molecular nitrogen (N$_2$) ice, while \citet{desch2021} proposed that impacts on extrasolar Pluto analogues would provide a plausible source for objects like `Oumuamua. However, \citet{levine2021} demonstrated that the necessary mass density of Pluto analogues for this formation to be plausible is unreasonably high. However, these issues are mitigated if `Oumuamua-like objects are produced exclusively by M-star systems and if impacts are generally at shallow angles \citep{Desch2022}.

In addition, many outgassing hypotheses fail to satisfactorily explain the lack of a dust coma. \citet{micheli2018} and \citet{seligman2021} argued that the postperihelion detection of `Oumuamua could explain the lack of dust, since subsurface cometary layers may be enriched with larger dust grains \citep{laufer2005,micheli2018} and `Oumuamua had likely already lost a fraction of its surface at detection \citep{SLB2019,SL2020,desch2021,jackson2021}. Processing in the interstellar medium, particularly when passing through supernova remnants, would also preferentially remove small dust grains from the surface of interstellar objects and {long-period comets} (LPCs)  via drag effects \citep{stern1987,stern1990}.

\citet{micheli2018} considered and rejected radiation pressure as the source of the anomalous acceleration, since this would require `Oumuamua to have bulk densities of $\rho<10^{-5}$ g cm$^{-3}$. However, \citet{bialyloeb2018}, \citet{moro-martin2019}, \citet{flekkoy2019}, \citet{luu2020}, and \citet{sekaninacomet} proposed various geometries and compositions which would allow radiation pressure to produce the observed non-Keplerian trajectory. \citet{bialyloeb2018} explored radiation pressure in the context of a millimeter-thick minor axis geometry, implying an artificial origin. However, radio signals from the object were not detected \citep[][{although this does not rule out activity}]{Enriquez2018,Tingay2018,Harp2019} and there is a $\sim$1\% probability for this geometry to produce the observed brightness variations \citep{Zhou2022}.

Separately, \citet{sekaninacomet} proposed a disintegrated dwarf comet as an origin. \citet{moro-martin2019}, \citet{flekkoy2019}, and \citet{luu2020} proposed that `Oumuamua could be a fractal, hyper-porous dust aggregate, formed either in a protoplanetary disk \citep{moro-martin2019} or in the inner coma of an Oort-cloud comet \citep{flekkoy2019,luu2020}. However, \citet{taylor2023} considered the effects of tidal forces on `Oumuamua, demonstrating that the dynamic viscosity of `Oumuamua needed to be relatively high to match the observed light curve. Other theories have invoked circumbinary systems \citep{Cuk2017,Jackson2017}, tidal fragmentation \citep{Raymond2018b,zhang2020tidal,Raymond2020}, and/or ejection from post--main sequence stars \citep{Hansen2017,Rafikov2018c,Katz2018}. Intriguingly, misaligned circumbinary systems are especially efficient progenitor systems of interstellar objects \citep{Childs2022}. 

\citet{Bergner2023} argued that radiolytic production of H$_2$ via cosmic ray exposure of amorphous H$_2$O ice could explain the nongravitational acceleration. This hypothesis may also explain the lack of dust produced from the object, because the H$_2$O bulk matrix is not sublimating.  For recent reviews concerning `Oumuamua and the second interstellar object 2I/Borisov, see \citet{Jewitt2022ARAA},  \citet{MoroMartin2022}, and \citet{Fitzsimmons2023}. 

However, questions still remain concerning `Oumuamua's outgassing and rotation state. \citet{flekkoy2019} reported a spin-down in `Oumuamua's rotational period during October 2017 observations. Separately, the constant torque imposed by \citet{Mashchenko2019} necessarily induces a spin-up in `Oumuamua's rotation. While this is in agreement with the known spin-up effect of outgassing \citep{Jewitt2003b,Drahus2011,Gicquel2012,Maquet2012,Fernandez2013,Wilson2017,Eisner2017,Roth2018,Kokotanekova2018,Biver2019,Combi2020,Jewitt2021,Jewitt22}, the model used would lead to `Oumuamua's destruction due to centrifugal forces. For example, \citet{Jewitt2019a,Jewitt2019b} predicted that the second interstellar object 2I/Borisov would suffer disintegration events due to the spin-up timescale implied by the nucleus size and production rates, which were later observed \citet{Jewitt2020}.

\citet{seligman2021} showed that an outgassing force directed from the substellar point --- although highly idealized --- would produce a complex rotation without the significant spin-up and disintegration predicted by \citet{rafikov2018b}. In this paper, we produce a model of small-body rotation using a substellar outgassing force, applying it to `Oumuamua. This model is capable of reproducing `Oumuamua's light curve and is stable over secular timescales. In addition, we demonstrate that under this outgassing model, the tumbling rotation of `Oumuamua is not an effect of `Oumuamua's close approach to the Sun, but is present before and after `Oumuamua's close approach to the solar system, despite damping of tumbling rotation.

This paper is organized as follows: in Section  \ref{sec:bigana}, we quantify tidal and outgassing torques on elongated small bodies via an idealized 2-dimensional model. Then in Section  \ref{sec:rotationmodel}, we define a model for the rotation state of elongated, ellipsoidal bodies similar to that defined in \citet{seligman2021}. This model ignores tidal forces, as justified by Section \ref{sec:bigana}{, but includes the torque from a Sun-induced substellar outgassing jet}. In Section \ref{sec:lightcurvefits}, we obtain a model which best reproduces the available photometric data for `Oumuamua. In Section \ref{sec:oumuamuafit}, we extend the model of `Oumuamua's rotational state both forward and backwards in time, finding the rotation state of `Oumuamua during its pre-- and post--solar system interstellar trajectories. In Section \ref{sec:discussion}, we {discuss our results, and in Section \ref{sec:conclusion}, we }conclude and discuss future research prospects. 

\begin{figure}
\centering
\includegraphics[width=\linewidth,angle=0]{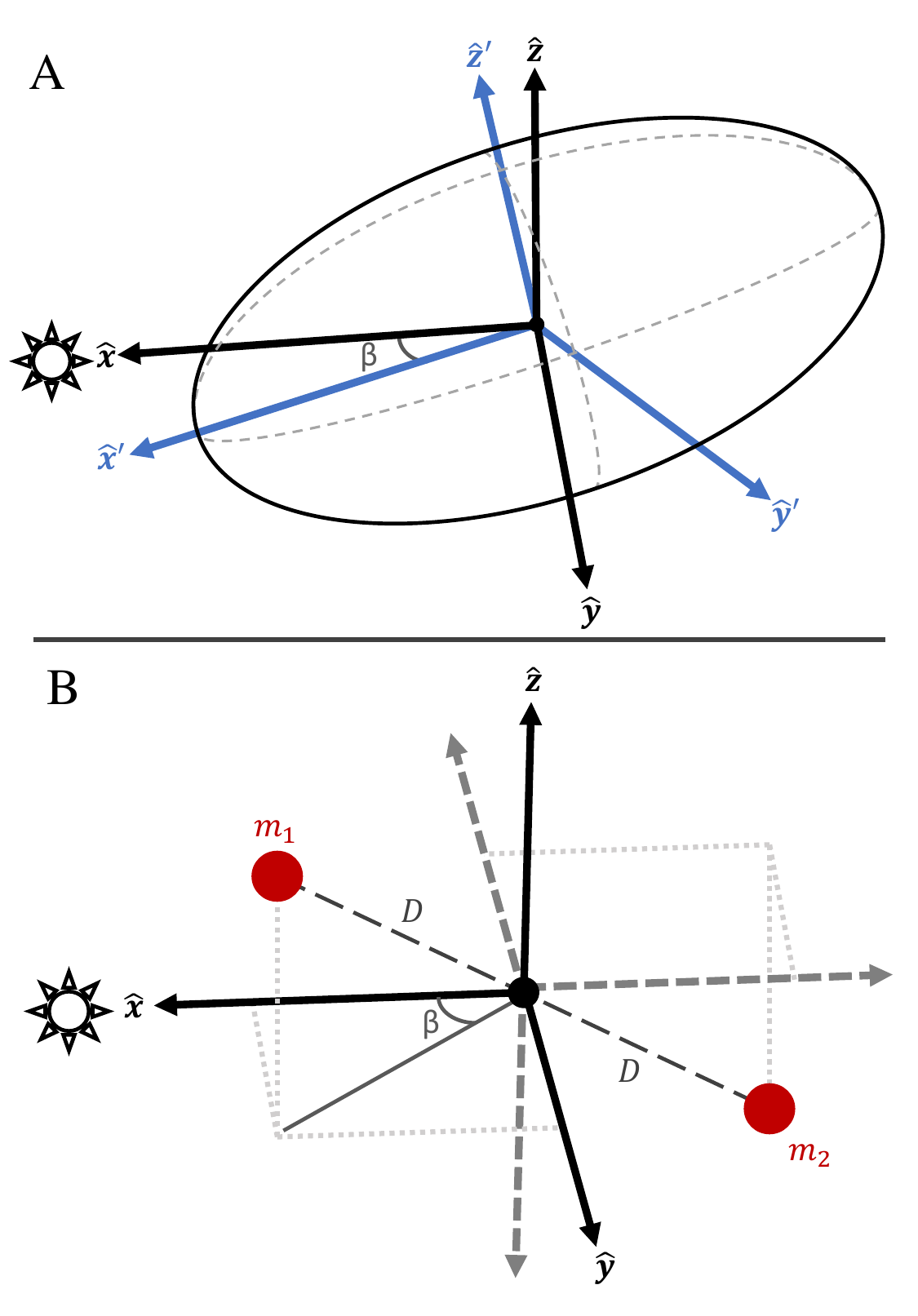}
\caption{Idealized models of an elongated ellipsoidal object. Panel A shows the ellipsoidal object, with the inertial (unprimed, black) coordinate system and the corotating (primed, blue) coordinate system. The object, and the corotating axes, are rotated by an angle $\beta$. Panel B shows the ``dumbbell" model characterized by two masses $m_1$ and $m_2$ at a distance $D$, again rotated by an angle $\beta$. Only the inertial coordinate system is shown. }
\label{fig:object_model}
\end{figure}

\section{Comparison of Outgassing and Tidal Torques}\label{sec:bigana}

In this section, we compare the effects of tidal- and outgassing-induced torques in order to justify neglecting tidal torques in the forthcoming rotation model. In Sections \ref{subsec:tidana} and \ref{subsec:jetana}, we calculate the magnitudes of each individual effect, which we compare in Section \ref{subsec:torquecomp}. Finally, in Section \ref{subsec:angleana}, we repeat this analysis with a more complex outgassing model. 

\subsection{Tidal Torques}\label{subsec:tidana}

In this subsection, we calculate the effect of tidal forces on the rotation of an ellipsoidal body with semi-axes $(a,b,c)$, $a=b,a>c$ and aspect ratio $\epsilon\equiv c/a$. We approximate the permanent quadrupole moment of the object as two concentric point masses of mass $m$ at a distance $D$ from the center of mass, similar to the `dumbbell' model developed by \citet{Batygin2015} and \citet{Seligman2021_spin}. The use of this model is validated in Appendix \ref{sec:dumbval}.

We find that for point masses located at $\pm (R,R,0)$, where $R\equiv\sqrt{a^2-c^2}$, the dumbbell model has a quadrupole moment $Q_D$ given by 
\begin{equation}\label{eq:dumbbellquad}
\begin{aligned}
    Q_D=\frac{4\pi}{15} {\rm diag}([&2x^2-y^2-z^2,\\&2y^2-x^2-z^2,\\&2z^2-x^2-y^2])\,.
\end{aligned}
\end{equation} 
This is equivalent to the quadrupole moment of an ellipsoid with semi-axes $a,b,c$. Schematic diagrams of this configuration are shown in Figure \ref{fig:object_model}. It is worth noting that for $a<c$, the point positions are distinct but the torques are equal. This situation is addressed in Appendix \ref{sec:cigtor}. 

We now provide a brief overview of the coordinate systems used in this {section}. The unprimed axes are inertial, and we set the x-axis in the direction of the Sun. The object rotates about the y-axis by an angle $\beta$ with respect to the x-axis. The primed axes corotate with the body and align with its three principal axes. 

The two point masses, denoted by $m_1$ and $m_2$, are located respectively at
\begin{equation}\label{eq:dumbbellpts}
    \boldsymbol{x'}_{\rm m_1/m_2}=\pm R\,\big(\cos\beta\,\boldsymbol{\hat{x}'}+\boldsymbol{\hat{y}'}-\sin\beta\,\boldsymbol{\hat{z}'}\big)\,.
\end{equation}
At a heliocentric distance $r_H$, the tidal acceleration is the difference in the acceleration between the center of mass and the point mass, and so the force is simply
\begin{equation}\label{eq:torqueforce}
    \boldsymbol{F}_{\text{tidal},m_1/m_2}=-\,G M_{\Sun}m \bigg(\,\frac{1}{\left(r_H\mp R\cos{\beta}\right)^2}-\frac{1}{r_H^2}\, \bigg)\,\boldsymbol{\hat{x}}\,.
\end{equation} 

The corresponding tidal torque is given by the cross product of Equations \eqref{eq:dumbbellpts} and \eqref{eq:torqueforce}. The vector component is $\mp R(\sin{\beta}\,\boldsymbol{\hat{y}}+\boldsymbol{\hat{z}})$, and we take only the y-component. The z-component of the torque arises from asymmetry across the x-axis in the `dumbbell' model not present in the ellipsoidal object and is therefore ignored. The magnitude of the total torque about the y-axis is then 
\begin{equation}\label{eq:tidal}
    \tau_{\text{tidal}}=\frac{8\pi}{15}G M_{\Sun} \left( \frac{r_Ha^5\rho\epsilon(1-\epsilon^2)\sin(2\beta)}{\big(r_H^2- a^2(1-\epsilon^2)\cos^2{\beta}\big)^2} \right),
\end{equation} 
where we have substituted $a=b$, $R=\sqrt{a^2-c^2}$, $m=(4\pi/15)abc\rho$, and $\epsilon=c/a$. Notably, this is equivalent to the y- or z-axis rotating torque for a body with $b=c$ and $a>c$ (see Appendix \ref{sec:cigtor}).

\subsection{Outgassing Torques}\label{subsec:jetana}

In this subsection, we calculate the magnitude of outgassing torques on cometary bodies. We assume this force operates at the substellar point and normal to the surface of the ellipsoidal body, as in \citet{SLB2019}.

In this model, we only consider the rotation about the y-axis, which is equivalent to x-axis rotation. The z-axis rotation does not evolve due to the assumed symmetry.

In the primed body frame, the equation of the ellipsoid is given by the relationship
\begin{equation}
    f_{\rm 3d}(x',y',z')=\frac{x'^2}{a^2}+\frac{y'^2}{b^2}+\frac{z'^2}{c^2}-1=0\,,
\end{equation}
and the unit vector normal to this surface is 
\begin{equation}\label{eq:norm}
    \frac{\nabla f_{\rm 3d}}{\|\nabla f_{\rm 3d}\|}=\left[\frac{x'^2}{a^4}+\frac{y'^2}{b^4}+\frac{z'^2}{c^4}\right]^{-1/2}\left[\frac{x'}{a^2}\boldsymbol{\hat{x}}+\frac{y'}{b^2}\boldsymbol{\hat{y}}+\frac{z'}{c^2}\boldsymbol{\hat{z}}\right]. 
\end{equation}
We further define a normalization factor, $n$, to be
\begin{equation}\label{eq:f_def}
    n\equiv\sqrt{(a\cos\beta)^2+(c\sin\beta)^2}\,.
\end{equation} 
Then the position of the substellar point, $(x_{ss}',y_{ss}',z_{ss}')$, is located where Equation \eqref{eq:norm} is equal to the Sun-pointing vector in the body frame. Although there is a sign degeneracy, we choose the negative variant to ensure that the outgassing jet points towards the Sun rather than away. Since $n$ is positive, the positions of the substellar point are given by: 
\begin{equation}\label{eq:ssprimepoints}
\begin{aligned}
\begin{dcases}
    x_{ss}'=-a^2\cos\beta/n\\
    y_{ss}'= 0 \\ 
    z_{ss}'=-c^2\sin\beta/n\,.
\end{dcases}
\end{aligned}
\end{equation}

The point can be rotated to the unprimed substellar point $(x_{ss},\,y_{ss},\,z_{ss})$ via the rotation matrix $\boldsymbol{R}$. The torque is then given by the cross product of vector pointing from the center of the body to the substellar point with the applied force vector. We assume that the outgassing acceleration scales with some constant $B$ such that the force due to outgassing is $B\, M\, r_H^{-2}$, where $r_H$ is the heliocentric distance and $M$ is the mass. We also assume that the object has a constant density $\rho$ and that the aspect ratio, $\epsilon=c/a$,  is constant for both small axes such that $a=b$. The magnitude of the outgassing torque which operates along the y-axis $\tau_{\text{jet}}$ is then finally given by
\begin{equation}\label{eq:jet}
\begin{aligned}
    \tau_{\text{jet}}=\,\bigg(\,\frac{2\pi}{3}Ba^4\rho\,\bigg)\,\bigg( \frac{\epsilon(1-\epsilon^2)\sin(2\beta)}{r_H^2\sqrt{\cos^2\beta+\epsilon^2\sin^2\beta}}\,\bigg)\,.
\end{aligned}
\end{equation}

\begin{figure}
    \centering
    \includegraphics[width=\linewidth,angle=0]{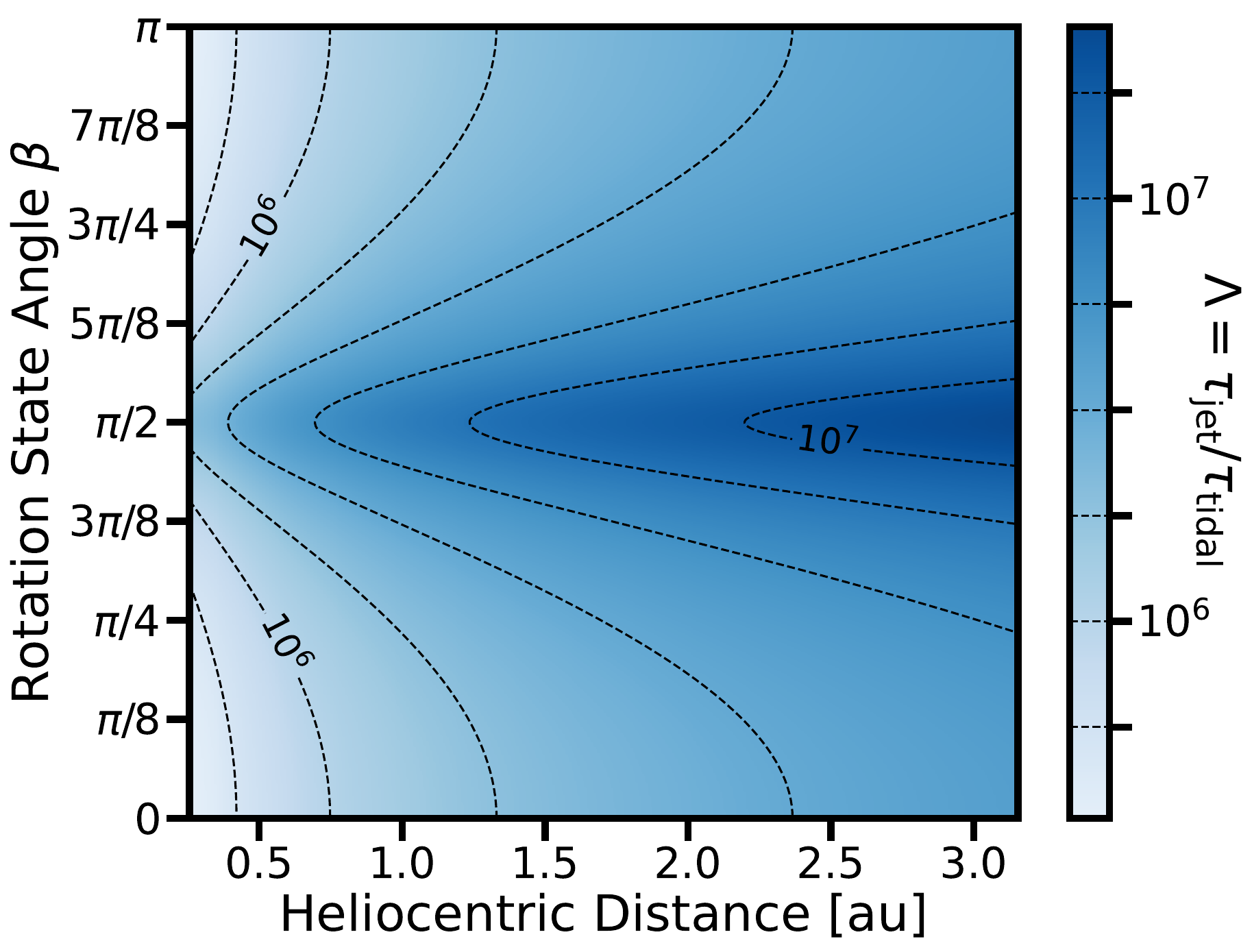}
    \caption{The ratio of outgassing to tidal torques for `Oumuamua, which is given by Equation \eqref{eq:torqueratio}. The ratio is plotted for a range of heliocentric distances and the rotation state of the body, which reflects the angle between the body's primary axis and the Sun. The contours are plotted every $10^{0.25}$, and are shown on the colorbar.}
    \label{fig:analytic_ratio}
\end{figure}

\subsection{Comparison of Outgassing and Tidal Torque}\label{subsec:torquecomp}

In this subsection, we compare the magnitudes of the outgassing and tidal torques. We define the ratio of the torques as $\Lambda\equiv\tau_{\mathrm{jet}}/\tau_{\mathrm{tidal}}$, which can also be written as
\begin{equation}\label{eq:torqueratio}
    \Lambda(a,c,r_H,\theta)=\frac{5B}{4GM_\Sun}\left(\frac{(r_H^2-(a^2-c^2)\cos^2{\beta})^2}{r_H^3\sqrt{(a\cos\beta)^2+(c\sin\beta)^2}}\right)\,.
\end{equation}
For `Oumuamua, we set $B=1.101\times 10^{23}$ cm$^3$ s$^{-2}$ \citep{micheli2018}, and $a:b:c=119:111:19$ m \citep{Mashchenko2019}. The ratio of outgassing to tidal torques for a range of heliocentric {distances} $r_H$ and rotation state angle $\beta$ is shown in Figure \ref{fig:analytic_ratio}. For these values, the minimum value of $\Lambda$ is at least $10^5$. Clearly, the outgassing torque dominates the rotational dynamics under these idealized assumptions. 

\begin{figure}
\centering
\includegraphics[width=\linewidth,angle=0]{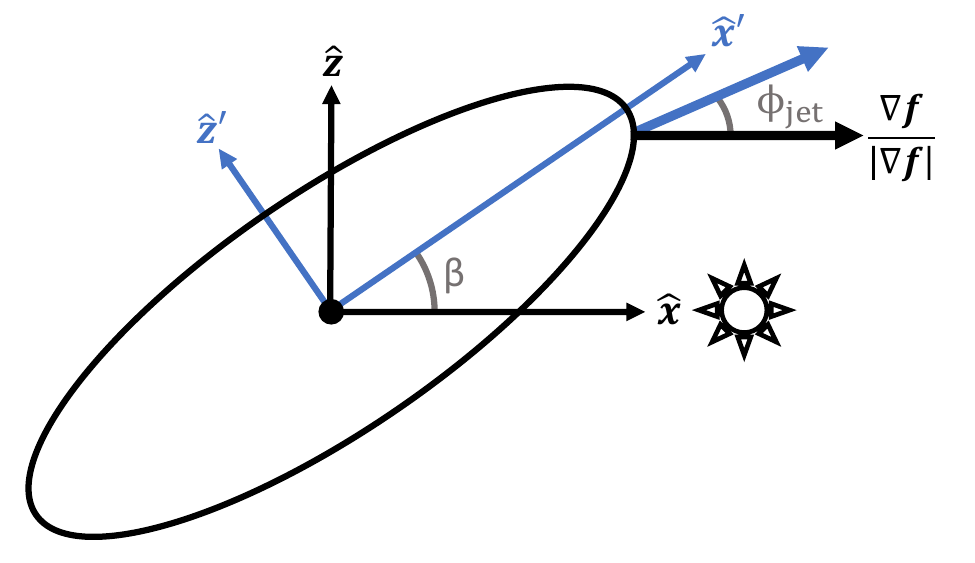}
\caption{Diagram showing the structure and defined angles of a nonnormal outgassing jet. The thick blue arrow indicates the outgassing jet direction, which is offset from the ellipsoid normal direction (the thick black arrow, given by Equation \eqref{eq:norm}) by the angle $\phi_{\rm jet}$. The corotating frame (thin blue arrows) is rotated with respect to the inertial frame (thin black arrows) by the angle $\beta$. }
\label{fig:outgas_rot}
\end{figure}

\subsection{Variations in Outgassing Angle}\label{subsec:angleana}

In this subsection, we calculate the outgassing torque with more general assumptions. Specifically, we relax the assumption that the outgassing vents normal to the surface, instead setting the direction of the outgassing to be offset from the normal by an angle $\phi_{\rm jet}\in[-\pi/2,\pi/2]$ (see Figure \ref{fig:outgas_rot} for a diagram). The force is then along the direction 
\begin{equation}
    \boldsymbol{\hat{F}}=\cos{\phi_{\rm jet}}\,\boldsymbol{\hat{x}}+\sin{\phi_{\rm jet}}\,\boldsymbol{\hat{z}}\,.
\end{equation} 

We take the cross product of the substellar point ($x'_{ss}$, $y'_{ss}$, $z'_{ss}$) with this force to obtain the torque. We find that the magnitude of the torque $\tau_{\text{rotjet}}$ is given by
\begin{equation}\label{eq:rotjettorque}
\begin{split}
    \tau_{\text{rotjet}}=\frac{2\pi}{3}\frac{\epsilon\rho a^4 B}{r_H^2\sqrt{\cos^2(\beta)+\epsilon^2\sin^2(\beta)}}\\
    \Big(\big((\epsilon^2+1)-(\epsilon^2-1)\cos2\beta\big)\sin\phi_{\rm jet}\\
    -(\epsilon^2-1)\sin2\beta\cos\phi_{\rm jet}\Big)\,.
\end{split}
\end{equation}

\begin{figure}
\centering
\includegraphics[width=\linewidth,angle=0]{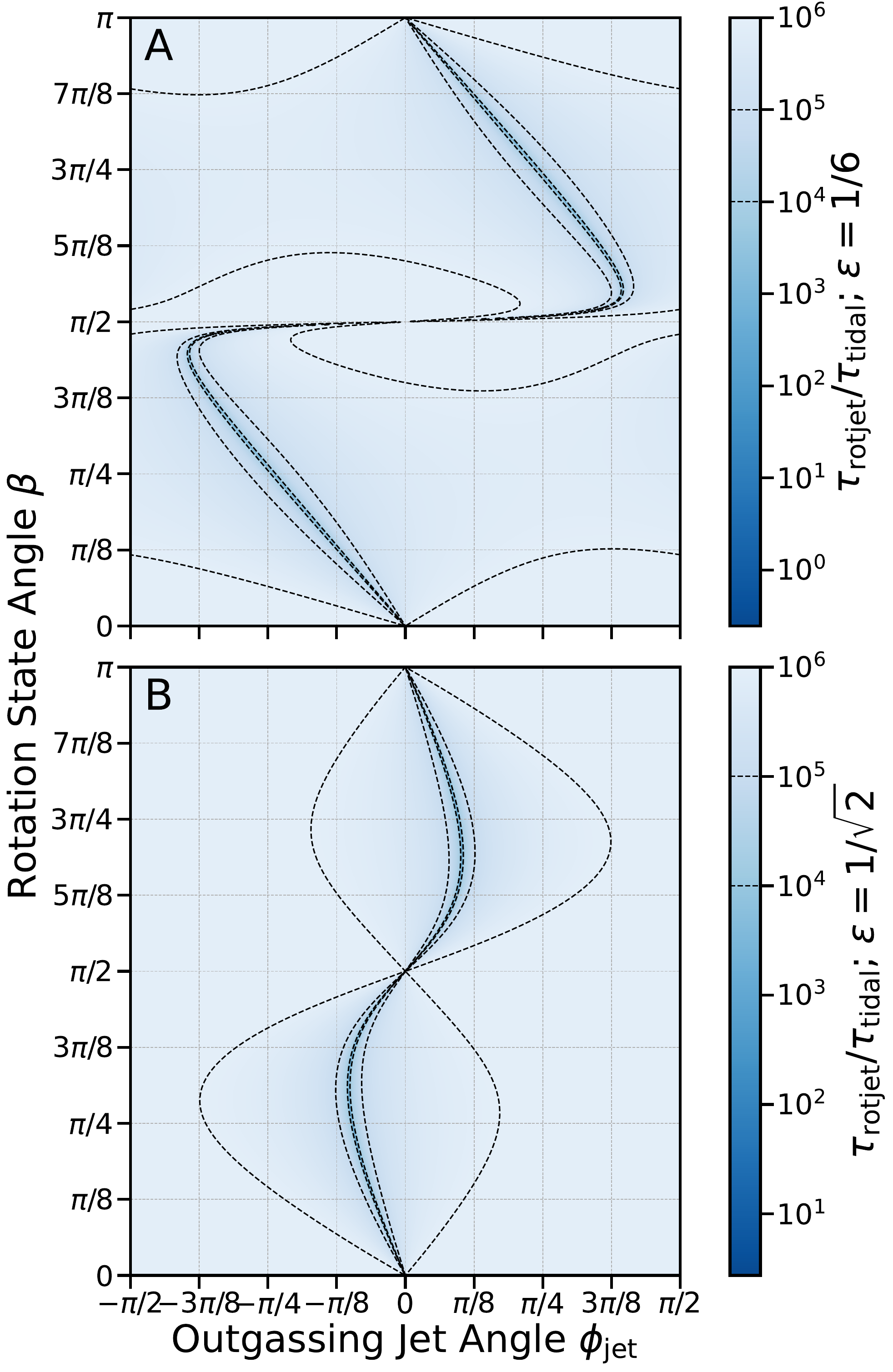}
\caption{The ratio of the nonnormal outgassing torque (Equation \eqref{eq:rotjettorque}) to the tidal torque (Equation \eqref{eq:tidal}), assuming that $a=115$ m and $r_H=0.256$ au. Panel A assumes that the aspect ratio $\epsilon$ is equal to $1/6$, characteristic for `Oumuamua. Panel B assumes that $\epsilon=1/\sqrt{2}$, typical for most small bodies. Contours are plotted at $10^4,\,10^5,\,10^6$ for both subplots.}
\label{fig:rot_ratio}
\end{figure}

In Figure \ref{fig:rot_ratio}, we show the magnitude of the ratio of the outgassing to tidal torque for a range of $\beta\in[0,\pi)$ and $\phi\in[-\pi/2,\pi/2]$, with $\epsilon=1/6,1/\sqrt{2}$ (characteristic of `Oumuamua and most minor bodies respectively) and $r_H=0.26$ au (`Oumuamua's perihelial distance). For $\epsilon=1/6$, the minimum value of the ratio is $\Lambda\simeq 1/4$. Therefore, while it is possible to construct a pathological scenario where both torques are comparable, the outgassing torque dominates the dynamics for most of the parameter space. It is therefore reasonable to ignore the effects of tidal torques for rotation models dealing with `Oumuamua.

\section{Ellipsoid Rotation Model}\label{sec:rotationmodel}

In this section, we present a dynamical rotation model for ellipsoidal bodies, which incorporates an outgassing jet directed from the substellar point. While this model does not incorporate the torque due to a tidal force, we demonstrated in Section \ref{sec:bigana} that the outgassing torque will dominate the tidal torque in all but the most pathological of scenarios (for the case of `Oumuamua). We begin by describing a model for the outgassing-driven rotation evolution, and then describe the model used to compute the synthetic light curve for a given rotation state.

\subsection{Rotation State Model}\label{subsec:rotmodel}

We consider the rotation of an ellipsoid with principal axes $a,\,b,\,c$. There exist two reference frames --- the inertial frame, {defined with respect to the Solar System,} and the (primed) corotating frame, where the $x',\,y',\,z'$ axes are aligned with the ellipsoid's principal axes. We will primarily represent and model the rotation state of the body in the corotating frame. We represent the rotation state with a quaternion $\boldsymbol{q}$, which represents the transformation from the corotating to the inertial frame (such that $\boldsymbol{x} = \boldsymbol{q}*\boldsymbol{x'}*\boldsymbol{q}^{-1}$). Quaternions are used for their speed and simplicity.

In the noninertial corotating frame, the torque equation 
can be solved for $\boldsymbol{\Dot{\omega}}$, which results in the following relationship,
\begin{equation}\label{eq:angacc}
    \boldsymbol{\Dot{\omega}} = \boldsymbol{J}^{-1}\boldsymbol{\tau}-\boldsymbol{J}^{-1}\big[\boldsymbol{\omega}\times(\boldsymbol{J}\boldsymbol{\omega})\big]\,.
\end{equation} 
In Equation \eqref{eq:angacc}, `$\times$' represents the 3-vector cross product, $\boldsymbol{J}$ represents the moment of inertia {tensor} (which is constant in the corotating frame), $\boldsymbol{\omega}$ is the corotating angular velocity vector, and $\boldsymbol{\Dot{\omega}}$ is the time derivative of the angular velocity.

The angular velocity $\boldsymbol{\omega}$ is related to the time derivative of the position quaternion. Since both are written in the corotating frame, $\boldsymbol{\Dot{q}}=(\boldsymbol{q}*\boldsymbol{\omega})/2$. We construct an integrator algorithm using the classic fourth-order Runge-Kutta scheme \citep{nummethods} for $\boldsymbol{q},\,\boldsymbol{\omega}$ simultaneously using the equations for both $\boldsymbol{\Dot{q}}$ and $\boldsymbol{\Dot{\omega}}$. This requires an expression for the torque at a given rotation state.

The torque is assumed to be exerted by an outgassing jet at the substellar point and normal to the surface. The point at which this jet originates (in the corotating frame) is solved by a generalization of the equations in Section \ref{subsec:jetana} to ellipsoidal bodies with a 3-dimensional NPA rotation. This can be solved by setting Equation \eqref{eq:norm} equal to the Sun-pointing vector in the corotating (primed) body frame, which we assume to be $x'_\Sun \boldsymbol{\hat{x'}}+y'_\Sun \boldsymbol{\hat{y'}}+z'_\Sun \boldsymbol{\hat{z'}}$, where $x'_\Sun,\,y'_\Sun,\,z'_\Sun$ can be calculated from the rotation state of the body. We generalize the (now 3-dimensional) normalization constant defined by Equation \eqref{eq:f_def} to be
\begin{equation}\label{eq:fnorm}
    n\equiv\sqrt{(a x'_\Sun)^2+(b y'_\Sun)^2+(c z'_\Sun)^2}\,.
\end{equation}
The coordinates of the substellar point in the body frame are therefore given by
\begin{equation}\label{eq:substellar}
\begin{dcases}
    x'_{ss}=-a^2 x'_\Sun /n \\
    y'_{ss}=-b^2 y'_\Sun /n \\
    z'_{ss}=-c^2 z'_\Sun/n
\end{dcases}\,.
\end{equation}
The normalized gradient vector (Equation \eqref{eq:norm}) is equivalent to the Sun-pointing vector defined by Equation \eqref{eq:substellar}. 

The force imparted by the outgassing jet is directed along the Sun-pointing vector and scaled by a magnitude $F(r_H)$, where $r_H$ is {heliocentric distance}. Therefore, the torque (in the body frame) due to this force is simply $\boldsymbol{\tau '}=\boldsymbol{r}'\times\boldsymbol{F}'$, where $\boldsymbol{r}'$ is the vector point described by Equation \eqref{eq:substellar} and $\boldsymbol{F}'$ is $F(r_H)\,[x'_\Sun \boldsymbol{\hat{x'}}+y'_\Sun \boldsymbol{\hat{y'}}+z'_\Sun \boldsymbol{\hat{z'}}]$, the Sun-directed force vector. Using the rotation quaternion $\boldsymbol{q}$, we can compute the coordinates of the Sun-pointing vector in the corotating frame. 

We now use this evolving rotation model to compute a synthetic light curve for the object when viewed from Earth. We construct and compare two models, one with geometric scattering \citep{Connelly1984} and one with a Lommel-Seeliger scattering surface \citep{muinonen2015}. The geometric scattering model is simplified, and describes a very general surface without multiple scattering or phase angle dependence. The Lommel-Seeliger model is a diffuse reflection model, again with isotropic single-scattering. The Lommel-Seeliger scattering law is commonly used for asteroids, since it is a {good} approximation for dark, low-albedo objects \citep{muinonen2015}. 

\subsection{Geometric Light Curve Model}\label{subsec:geomag}

 In this subsection, we summarize and describe a model for computing a synthetic light curve for an ellipsoid with geometric scattering, as described by \citet{Connelly1984}. In the geometric scattering model, the brightness is assumed to scale with the projected visible, illuminated area. In the corotating frame of the ellipsoid, the $\boldsymbol{\hat{e}}_\Sun$ and $\boldsymbol{\hat{e}}_\Earth$ are the unit Sun- and Earth-pointing vectors, respectively. We also define the matrix $\boldsymbol{C}$ as 
 \begin{equation}
     \boldsymbol{C}\equiv{\rm diag}([a^{-2},\,b^{-2},\,c^{-2}])\,.
 \end{equation}
The projected area, $A$, can be written as
\begin{equation}\label{eq:geobrightarea}
    A=\pi\,a\,b\,c\frac{(\boldsymbol{\hat{e}}_\Earth^T\boldsymbol{C}\boldsymbol{\hat{e}}_\Earth)^{1/2}\,(\boldsymbol{\hat{e}}_\Sun^T \boldsymbol{C}\boldsymbol{\hat{e}}_\Sun)^{1/2}\,+\,\boldsymbol{\hat{e}}_\Earth^T\boldsymbol{C}\boldsymbol{\hat{e}}_\Sun}{2(\boldsymbol{\hat{e}}_\Sun^T\boldsymbol{C}\boldsymbol{\hat{e}}_\Sun)^{1/2}}\,.
\end{equation}
The derivation of this relationship is beyond the scope of this paper, but can be found in Section 3 of \citet{Connelly1984}. Since the brightness scales with this projected area, the magnitude of the object is equal to 
\begin{equation}\label{eq:COmodel}
    H=\Delta M - 2.5\log(A)\,.
\end{equation}
In Equation \eqref{eq:COmodel}, $\Delta M$ corrects for both the conversion to magnitudes and for the brightness-area scaling constant{, which includes the albedo}. Since it is relatively simple to obtain a $\Delta M$ which effectively matches a light curve, this model is an efficient way to generate a synthetic light curve for an ellipsoidal body. 

\subsection{Lommel-Seeliger Light Curve Model}\label{subsec:MLmag}

In this subsection, we describe a model which computes a synthetic light curve for an ellipsoid with a Lommel-Seeliger scattering surface, which is described in \citet{muinonen2015}. For a Lommel-Seeliger scattering surface, the incoming radiation is assumed to decrease exponentially with optical depth, with diffusive isotropic single-scattering from the surface. \citet{muinonen2015} used this model to derive a formula for the light curve of a rotating ellipsoid for a given astrometric arrangement. This numerical model is summarized below. 

We assume an isotropic single-scattering phase function $P(\psi)=1$. This incorporates modulation based on the body orientation, but assumes no additional modulation from the scattering function. We define the phase angle $\psi$ as the interior Sun-object-Earth angle, and note that $\cos\psi=\boldsymbol{\hat{e}}_{\Sun}\cdot\boldsymbol{\hat{e}}_{\Earth}$, where $\boldsymbol{\hat{e}}_{\Sun}$ and $\boldsymbol{\hat{e}}_\Earth$ are defined as in Section \ref{subsec:geomag}. We also define the matrix $\boldsymbol{C}$ as in that Section. Now, the parameters $T_{\Sun}$ and $T_\Earth$ {are defined} to be
\begin{equation}
    \begin{aligned}
        T_{\Sun}\equiv&\sqrt{\boldsymbol{\hat{e}}_{\Sun}^TC\boldsymbol{\hat{e}}_\Sun}\\
        T_\Earth\equiv&\sqrt{\boldsymbol{\hat{e}}_\Earth^TC\boldsymbol{\hat{e}}_{\Earth}}\,.
    \end{aligned}
\end{equation}
We further define the parameter $T$ as
\begin{equation}
    T\equiv\sqrt{T_{\Sun}^2+T_\Earth^2+2T_{\Sun}T_\Earth\cos\psi'}\,,
\end{equation}
the angle $\psi'$ as 
\begin{equation}
\begin{aligned}
    \cos\psi'=&\frac{\boldsymbol{\hat{e}}_{\Sun}^TC\boldsymbol{\hat{e}}_\Earth}{T_{\Sun}T_\Earth}\,,
\end{aligned}
\end{equation}
and the angle $\lambda'$ as 
\begin{equation}
\begin{aligned}
    \cos\lambda'=&\frac{T_\Sun+T_\Earth\cos\psi'}{T}\\
    \sin\lambda'=&\frac{T_\Sun\sin\psi'}{T}\,.
\end{aligned}
\end{equation}
Then the absolute magnitude is given by
\begin{equation}\label{eq:MLmodel}
\begin{aligned}
        H=\Delta M-&2.5\log\bigg(abc\frac{T_\Sun T_\Earth}{T}\\
        &\big(\cos(\lambda'-\psi')+\cos\lambda'+\sin\lambda'\sin(\lambda'-\psi')\\
        &\ln{\big[\cot(\frac{1}{2}\lambda')\cot(\frac{1}{2}(\psi'-\lambda'))\big]}\big)\bigg),
\end{aligned}
\end{equation}
where $\Delta M$ is once again a constant to absorb the multiplicative constants {(such as the albedo)} and magnitude conversion, as in Section \ref{subsec:geomag} and \cite{Mashchenko2019}.
 
\section{Light Curve Fitting}\label{sec:lightcurvefits}

In this section, we describe the process which we use to obtain an optimal rotation model from photometric data. This methodology is agnostic of object and scattering model. Note that this algorithm does not {have a convergence criterion}, and therefore extensive CPU time is required to find an optimal parameter value.

\subsection{Parameterization}\label{subsec:params}

First, we parameterize the rotation model described in Section \ref{sec:rotationmodel}. {The Sun- and Earth-pointing vectors, in the inertial frame, are defined to be $\boldsymbol{\hat{e}}_\Sun$ and $\boldsymbol{\hat{e}}_\Earth$, respectively. These vectors can be obtained from astrometric data, in this case via a numerical integration of the trajectory. }

Next we define parameters to characterize the {initial }rotation state. {Here, we diverge from the methodology of \citet{Mashchenko2019} and use a simple standard spherical angle model rather than Euler angles. We do this because i) this system is more intuitive than Euler angles and ii) the rotation model is directly dependent on these parameters, enabling more efficient optimization. However, these coordinate systems are interchangeable, and we present the conversion from our system to the Euler angles used by \citet{Mashchenko2019} in Section \ref{subsec:eulerangles}.} Since the magnitude of the outgassing force is already given by astrometric measurements (up to a factor of outgassing anisotropy), the time evolution of the rotation state is dynamically determined down to numerical noise. We therefore only need to set the initial state of the object. We define 6 parameters for this purpose --- 3 for the initial rotation of the inertial frame with respect to the corotating frame, and 3 for the initial angular {velocity} in the corotating frame. 

We characterize the initial rotation with 3 angles $\alpha,\,\beta,\,\gamma$, which are sufficient to define the rotation. While this structure necessarily allows for the issue of degree-of-freedom loss for certain parameter values, the parameter space degeneracy caused by this issue is quite small, and is avoided in the rotation model itself.
{To determine the orientation of the body}, we define a spherical coordinate system with a {polar} direction along the x-axis of the corotating frame, {which is aligned with the longest axis of the body}.\footnote{{Note that we work primarily in the corotating frame, and so this rotation defines the inertial frame relative to the corotating one.}} The $\alpha,\,\beta$ parameters specify an axis, with $\alpha$ the polar angle and $\beta$ the azimuthal angle. The final angle, $\gamma$, specifies an anticlockwise rotation about this axis. We can then write an initial rotation quaternion as 
\begin{equation}
\begin{aligned}
    \boldsymbol{q}_0=\cos(\gamma/2)+\sin(\gamma/2)\cos(\alpha)\,&\mathbf{i}\\
    +\sin(\gamma/2)\sin(\alpha)\cos(\beta)\,&\mathbf{j}\\
    +\sin(\gamma/2)\sin(\alpha)\sin(\beta)\,&\mathbf{k}\,.
\end{aligned}
\end{equation}

For the initial angular velocity, we define two angles $\theta,\,\phi$ and a {magnitude $\nu$}. As before, these two angles specify an axis, now with $\theta$ the polar angle and $\phi$ the azimuthal angle. The initial angular velocity $\boldsymbol{\omega}_0=\boldsymbol{J}^{-1}\boldsymbol{L}$ is along the direction determined by the angles and has a magnitude $\nu$. {All of these parameters are defined at UT 2017-10-25 01:04:16, the epoch of first observation.}

{Bounds for these parameter values are simple to determine.} For these 6 parameters, the polar angles $\alpha,\,\theta\in[0,\,\pi]$, the azimuthal angles $\beta,\,\phi\in[-\pi,\,\pi)$, and the additional rotation angle $\gamma\in[-\pi,\,\pi)$. While we set the angular frequency $\nu\geq0$ to restrict the degeneracy, the upper bound is theoretically infinite. In practice, it is set such that the object's rotation is not unreasonably fast, depending on the observed period of the object's light curve.

\begin{figure*}[p]
\begin{interactive}{animation}{oumuamua-anim.mp4}
\includegraphics[width=0.7\textwidth,angle=0]{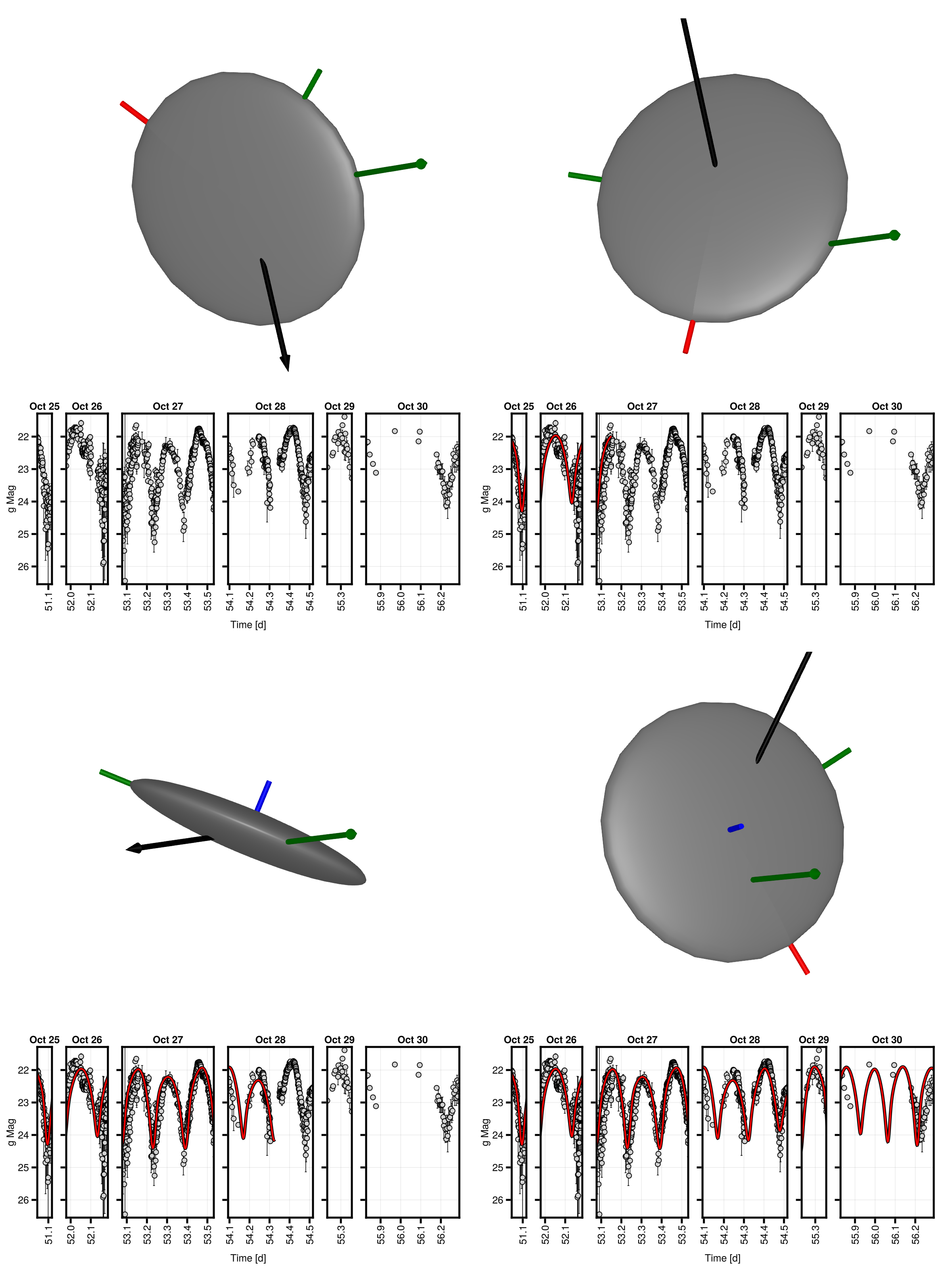}
\end{interactive}
\centering
\caption{An animation of the best-fitting rotation model for `Oumuamua, using the CO84 model and parameters given by Table \ref{table:optimalparams}. The upper panel shows a 3-dimensional model of an ellipsoid with dimensions of 115:111:19 m, in the rotation state at a given time. The red, green, and blue rods are aligned with these axes of the ellipsoid. The black arrow is {proportional to the angular momentum}. The green arrow is aligned with the substellar outgassing jet and is rendered with a nonphysical constant magnitude pointing towards the Sun. The camera is set such that the ellipsoid is viewed as if from the Earth. The lower row shows the photometric data (gray points) and the synthetic light curve (red line, {computed using the CO84 model}), {similar to the top row in Figure \ref{fig:optmodelfits}}. The synthetic light curve evolves with the rotating object. The animation progresses at $\sim$1.5 hr s$^{-1}$. { A grid of 4 still frames is shown in the print version corresponding to  times 51.045 (initial observation), 53.15 (light curve maximum), 54.3 (light curve minimum), and 56.3 (final October 2017 observation). All times are shown in modified Julian days with a zero point at 58000 MJD. In the animation, the ellipsoid is shown exhibiting NPA rotation, with the light curve appearing on the bottom row as described by the scattering model.}}
\label{fig:anim}
\end{figure*}

We finally add one additional parameter, a scaling constant $f$ with $f\in[0,\,1]$. This is used to account for the physically unrealistic model used to approximate the force due to outgassing --- namely, that it forms a single collimated jet originating from the substellar point. This assumption is relatively restrictive, since large swaths of the object's surface may be irradiated and outgassing, and these additional jets will simultaneously contribute to the observed astrometric acceleration and counteract the torque induced by the substellar outgassing jet. Rather than use a more complex outgassing model, which would significantly worsen the efficiency of the optimization, we elect to instead scale the outgassing acceleration by $f$, and additionally optimizing over this parameter. This constant essentially characterizes the dependence of the outgassing acceleration on the solar irradiance, and approximates a more physically-accurate full-surface model. {We note that a fully physical model must necessarily incorporate temperature-dependent sublimation and gas escape from the cometary matrix, which are not well-determined.}

\subsection{{Conversion to Euler Angles}}\label{subsec:eulerangles}

{The free parameters used in \citet{Mashchenko2019} are distinct from the ones used in this work. However, for convenience of comparison, we convert the free parameters used here into those used in that work and present them alongside our parameters (Table \ref{table:optimalparams}). \citet{Mashchenko2019} uses 6 parameters for the initial rotation state --- the angular momentum magnitude $L$, the angular momentum polar angle $\theta_L$, the angular momentum azimuthal angle $\varphi_L$, the initial reduced kinetic energy $E'$, the precession Euler angle $\varphi_0$, and the rotation Euler angle $\psi_0$. }

{Conversion between the parameters for the angular velocity and angular momentum are straightforward because $\boldsymbol{L}=\boldsymbol{J}\boldsymbol{\omega}$, and $\boldsymbol{L}=L\cos\theta_L\boldsymbol{\hat{x}}+L\sin\theta_L\cos\varphi_L\boldsymbol{\hat{y}}+L\sin\theta_L\sin\varphi_L\boldsymbol{\hat{z}}$. The Euler angles are related to a rotation state quaternion $\boldsymbol{q}=s+a\boldsymbol{i}+b\boldsymbol{j}+c\boldsymbol{k}$ by
\begin{subequations}\label{eq:eulerangles}
\begin{align}
    \varphi=&\arctan\Big(\frac{2(sa+bc)}{1-2(a^2+b^2)}\Big)\,,\\
    \psi =&\arctan\Big(\frac{2(sc+ab)}{1-2(b^2+c^2)}\Big)\,,\\
    \theta=&\arcsin\Big(2(sb-ac)\Big)\,.\label{eq:eulertheta}
\end{align}
\end{subequations}
In Equation \eqref{eq:eulertheta}, $\theta$ is the nutation angle, which is not a free parameter in \citet{Mashchenko2019}. Instead, that text uses the reduced rotational kinetic energy $E'$, which can be computed from 
\begin{equation}\label{eq:rotkinetic}
    E'=1+\sin^2\theta\,\big(\sin^2\psi\,(\boldsymbol{J}_b^{-1}-\boldsymbol{J}_c^{-1})+\boldsymbol{J}_c^{-1}-1\big)\,.
\end{equation}
In Equation \eqref{eq:rotkinetic}, $\boldsymbol{J}$ is the moment of inertia, and the subscripts indicate axial components.}

{It is critical to note, however, that \citet{Mashchenko2019} defines parameters relative to the inertial frame, while we define ours relative to the corotating frame. As a result, conversion from our parameters requires  computation of the initial rotation state quaternion and angular velocity, conversion  to the inertial frame, and then application of the definition of the angular momentum and Equations \eqref{eq:eulerangles} and \eqref{eq:rotkinetic} to find the relevant parameters. }

\begin{figure*}
\centering
\includegraphics[width=\linewidth,angle=0]{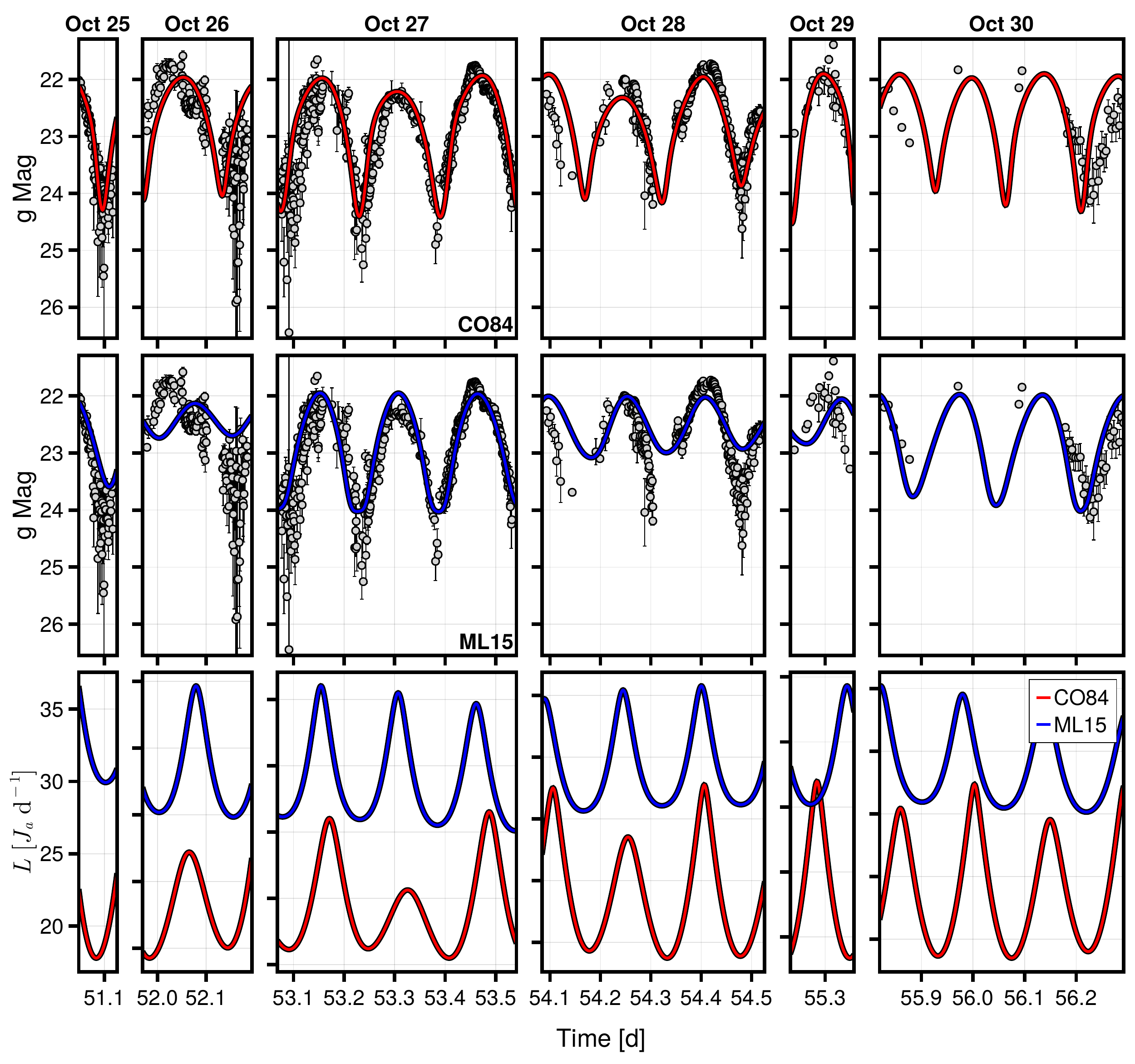}
\caption{Best-fit synthetic light curves (red lines) and photometric data (grey points). The data shown  are those taken in October 2017 only. The first row shows the CO84 model (Equation \eqref{eq:COmodel}), and the second shows the ML15 model (Equation \eqref{eq:MLmodel}). The parameters for each model are given in Table \ref{table:optimalparams}. The third row shows the instantaneous {normalized angular momentum magnitude $L$} of both models over the same time period. It is important to note that the modulation in the {angular momentum magnitude $L$} over the course of the observations is a result of the continuous application of outgassing torques, {because it is constant in the absence of torques}. {The time shown has units of days, with a zero point at 58000 MJD.}}
\label{fig:optmodelfits}
\end{figure*}

\subsection{Optimization Algorithm}\label{subsec:optalg}

We define the `optimal' rotation model to have the maximum likelihood for the given photometric data. In order to reduce the variation in the likelihood over the parameters, however, we will instead use the log-likelihood, which is comaximal with the likelihood. Assuming that the photometric data magnitudes are Gaussian-distributed, we write the log-likelihood as 
\begin{equation}\label{eq:lnlikelihood}
    \ln\mathcal{L}=-\sum_i\frac{(m_i-d_i)^2}{2\sigma_i^2}\,.
\end{equation}
In Equation \eqref{eq:lnlikelihood}, the $m_i$ are the synthetic magnitudes for a given time, the $d_i$ are the observed data for a given time, and the $\sigma_i$ are the uncertainties in those measurements. 

Recall, however, that our synthetic data are calculated down to an arbitrary additive constant. Fortunately, it is simple to analytically compute an additive constant which maximizes the log-likelihood. By writing $m_i\rightarrow m_i+C$, and taking $\partial\ln\mathcal{L}/\partial C=0$, we find that 
\begin{equation}
    C = -\bigg(\sum_i\frac{m_i-d_i}{\sigma_i^2}\bigg)\bigg(\sum_i\frac{1}{\sigma_i^2}\bigg)^{-1}\,.
\end{equation}
In order to optimize this function, we use the Multi-Level Single-Linkage (MLSL) algorithm \citep{RinnooyKan1987}, modified with a Sobol low-discrepancy sequence \citep{Kucherenko2005} to perform a global search. Local searches are accomplished by the use of the Bound Optimization BY Quadratic Approximation (BOBYQA) algorithm \citep{Powell2009}, with relative tolerances for the parameter and function value set to $10^{-4}$ to increase efficiency. The optimal parameter set found in the MLSL search is then further locally optimized by the BOBYQA algorithm with relative tolerances of $10^{-15}$, refining the answer. The MLSL convergence criterion is difficult to reach, leading to an impractically-long wall time. To adjust for this, we set a maximum number of evaluations in the algorithm. While we cannot confirm that the result found is a global optimum, we verify by inspection that the model sufficiently reproduces the photometric data. 

In this paper, we implement these algorithms in Julia 1.8.5 \citep{Julia}. The optimization algorithms are applied from the NLopt library \citep{NLopt} in Julia. 

\section{`Oumuamua Rotation Model}\label{sec:oumuamuafit}

In this section, we present an optimal rotation model for `Oumuamua, which was found using the algorithms described in Sections \ref{sec:rotationmodel} and \ref{sec:lightcurvefits}. 

\subsection{Fitting Rotation Parameters}\label{subsec:oumuamuafit}

Photometric data exist for `Oumuamua for the dates of October 25-30 2017 and November 21-23 2017, collectively published in \citet{belton2018}. The data in November 2017 are notably sparse, with only 48 data points over a 3-day period, limiting accuracy for a significant delay in optimization (as November 2017 fitting requires simulation over an additional month). Therefore, we only fit our model to the data taken in October 2017. This simplification  removes the necessity of numerically integrating the torque model for the intervening month and therefore significantly improves the processing time of our optimization. 

\begin{table}[ht]
\centering
\caption{The optimal initial rotation parameters for `Oumuamua, {with the parameters computed in this work and those used in \citet{Mashchenko2019}}. The $1\sigma$ uncertainty expressed for {our} parameters is given by the Cramér-Rao bound, which is obtained by fitting a Gaussian likelihood model to the region around the maximum. {The $1\sigma$ uncertainty in the Euler angles is computed by drawing parameter values from a corresponding distribution, converting them to the \citet{Mashchenko2019} variety by the method described in Section \ref{subsec:eulerangles}, and taking the standard deviation.} The given frequency $\nu$ is the instantaneous rotation frequency at the time of first observation, which corresponds to the reciprocal of the period. {To enable comparison to other models, we also report the reduced $\chi^2$ values for each model fit. Note that the reduced $\chi^2$ value and the log-likelihood are co-optimal. }}
\begin{tabular}{ crr } 
 Parameter: & \multicolumn{1}{c}{CO84:} & \multicolumn{1}{c}{ML15:} \\ 
 \hline\hline
 $\alpha$ {[rad]} & {$-0.8112\pm4.9\times10^{-5}$} & {$-0.5972\pm4.8\times10^{-5}$} \\
 $\beta$ {[rad]} & {$-1.7730\pm3.6\times10^{-5}$} & {$-1.4626\pm1.3\times10^{-4}$} \\
 $\gamma$ {[rad]} & {$\phantom{-}2.0782\pm4.9\times10^{-5}$} & {$\phantom{-}3.1414\pm5.6\times10^{-4}$} \\
 $\theta$ {[rad]} & {$\phantom{-}2.5319\pm5.6\times10^{-4}$} & {$\phantom{-}1.5930\pm1.2\times10^{-4}$} \\
 $\phi$ {[rad]} & {$\phantom{-}3.3884\pm2.3\times10^{-4}$} & {$\phantom{-}2.8937\pm2.0\times10^{-5}$} \\
 $\nu$ [hr$^{-1}$] & {$\phantom{-}0.1423\pm2.5\times10^{-6}$} & {$\phantom{-}0.2110\pm4.2\times10^{-5}$} \\
 $f$ & {$\phantom{-}0.5320\pm4.2\times10^{-5}$} & {$\phantom{-}0.6838\pm3.7\times10^{-5}$} \\\hline
 {$L$ [$J_a$ d$^{-1}$]} & {$22.5642\pm1.9\times10^{-3}$} & {$36.5990\pm7.4\times10^{-3}$}\\
 {$\theta_L$ [rad]} & {$2.8791\pm5.3\times10^{-4}$} & {$1.6886\pm4.4\times10^{-4}$} \\
 {$\varphi_L$ [rad]} & {$4.1766\pm1.1\times10^{-3}$} & {$1.0654\pm3.6\times10^{-4}$} \\
 {$\varphi_0$ [rad]} & {$5.2445\pm1.6\times10^{-4}$} & {$3.3204\pm1.2\times10^{-3}$} \\
 {$\psi_0$ [rad]} & {$5.146\pm1.6\times10^{-4}$} & {$6.0167\pm8.5\times10^{-4}$} \\
 {$E'$ [$J_a^{-1}$]} & {$0.8951\pm3.8\times10^{-5}$} & {$0.5954\pm1.9\times10^{-4}$} \\\hline
 {$\chi^2$} & \multicolumn{1}{c}{{22.0418}} & \multicolumn{1}{c}{{25.0963}}
\end{tabular}
\label{table:optimalparams}
\end{table}

In our integration, we use a time step of $\Delta t = 0.001$ d. The outgassing acceleration, which is used as described in Section \ref{sec:rotationmodel}, is $4.92\times10^{-4}\, (r/{\rm 1\, au})^{-2}$ cm s$^{-2}$, or $3.67\times 10^4\, (r/{\rm 1\, au})^{-2}$ m d$^{-2}$. We set an upper bound for the $\nu$ values of {0.25 hr$^{-1}$}, which corresponds to a full rotation period of 4 hr {(computed as $1/\nu$)}. This value is below all estimates made of the rotation period by other authors \citep{drahus2017, meech2017, belton2018, fraser2018}, and more rapid rotation periods result in numerical instability in our torque model. Our optimization therefore allows for an instantaneous initial period of $P>4$ hr.

Over the course of the integration, the heliocentric distance $r$ {is} computed using linear interpolation between data obtained from the Jet Propulsion Laboratory's Horizons database\footnote{https://ssd.jpl.nasa.gov/horizons.cgi} with a half-hour cadence. {The Sun- and Earth-pointing vectors, $\boldsymbol{\hat{e}}_\Sun$ and $\boldsymbol{\hat{e}}_\Earth$, are computed using REBOUND \citep{rebound}.} `Oumuamua is assumed to be an ellipsoid with dimensions of 115:111:19 m, as found by \citet{Mashchenko2019} (assuming an albedo $p\simeq0.1$). {These values are also used to compute the moment of inertia $\boldsymbol{J}$, assuming a constant density.} Our model does not interpolate the integrated rotation values --- instead, we compute the rotation for every time at which data was taken, with additional times added such that each time step $\Delta t_i\leq0.001$ d. This results in approximately 6,000 time values for which the model is computed. Due to the efficiency of Julia, running this integration model for these 6,000 {timesteps} has a wall time of approximately 80 milliseconds,\footnote{{On a Dell Inc. Inspiron 14 Plus 7420 with a 12th Gen Intel i7 Core at 1.69 TFlops.}} enabling us to perform this high-performance optimization on a tractable timescale. {Since this algorithm has no convergence criterion}, we set a maximum of $10^6$ local optimizations, which sufficiently explores the parameter space and produces a result which, by inspection, sufficiently reproduces the photometric data. We compute optimal parameters for both the geometric scattering law of \citet{Connelly1984} (CO84) and the Lommel-Seeliger scattering law of \citet{muinonen2015} (ML15). An animation of the evolution of the optimal CO84 model is given in Figure \ref{fig:anim}. {To confirm the necessity of torque in this model, we also compute an optimal rotation state for a zero-torque model using the CO84 scattering law, which is given in Appendix \ref{sec:zerotorquemodel}.} 

The results of the optimization are given in Table \ref{table:optimalparams}. The uncertainties in these values are given by the Cramér-Rao bound \citep{Rao1992,Cramer1999}, which provides a lower limit on the parameter uncertainty of a maximum likelihood estimation. The Cramér-Rao uncertainty is given by $I(\boldsymbol{X})^{-1/2}$, where $I(\boldsymbol{X})$ is the Fisher information at the maximum likelihood estimate $\boldsymbol{X}$. For our purposes, the Fisher information for a single parameter $x_i$ is given by $-\partial^2 \ln{\mathcal{L}}(\boldsymbol{X})/\partial x_i^2 $. For our multivariate system, this is simply extended into a vector. We numerically estimate these values for each parameter using the second-order finite difference method. This method of estimating the uncertainty is essentially equivalent to fitting a Gaussian distribution to the likelihood in the region surrounding the optimum value. The synthetic light curve models corresponding to these optimal parameters are plotted versus the October 2017 photometric data along with the periods over time in Figure \ref{fig:optmodelfits}.

While the CO84 model produces a better fit to the photometric data than the ML15 model {(see Table \ref{table:optimalparams})}, the distinction is likely not statistically significant.

\subsection{Rotation Model Extensions}\label{subsec:projections}

\begin{figure}
\centering
\includegraphics[width=\linewidth,angle=0]{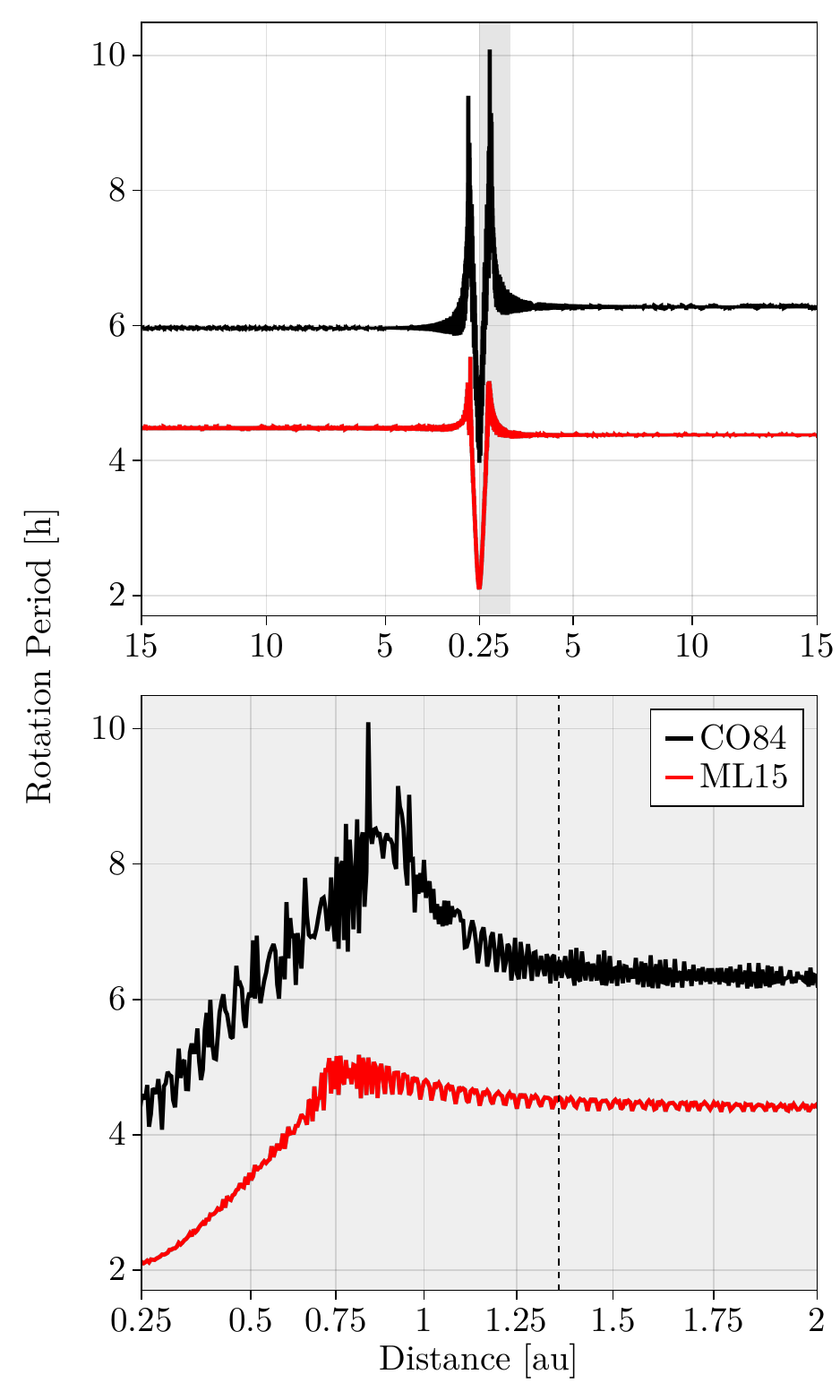}
\caption{Rotation period for `Oumuamua over its trajectory, plotted versus heliocentric distance. Both optimal parameter sets (Table \ref{table:optimalparams}) are shown. Because the time scale of the trajectory ($\sim$5 yr) is significantly longer than the $\sim1$ hour-scale modulation effect of the outgassing torques, we present the daily average of the period rather than its instantaneous value. We show the rotation state over the entire trajectory within 15 au inbound and outbound (top panel) and a zoom-in between perihelia and when the \textit{Spitzer} observations occurred at $\sim$2 au (bottom panel). The zoom-in is shown on the top panel by a gray highlight. The first photometric observations were obtained when the object was at $\sim$1.35 au, denoted with a vertical dashed line in the second panel. }
\label{fig:angmommag}
\end{figure}

\begin{figure*}
\centering
\includegraphics[width=0.88\linewidth,angle=0]{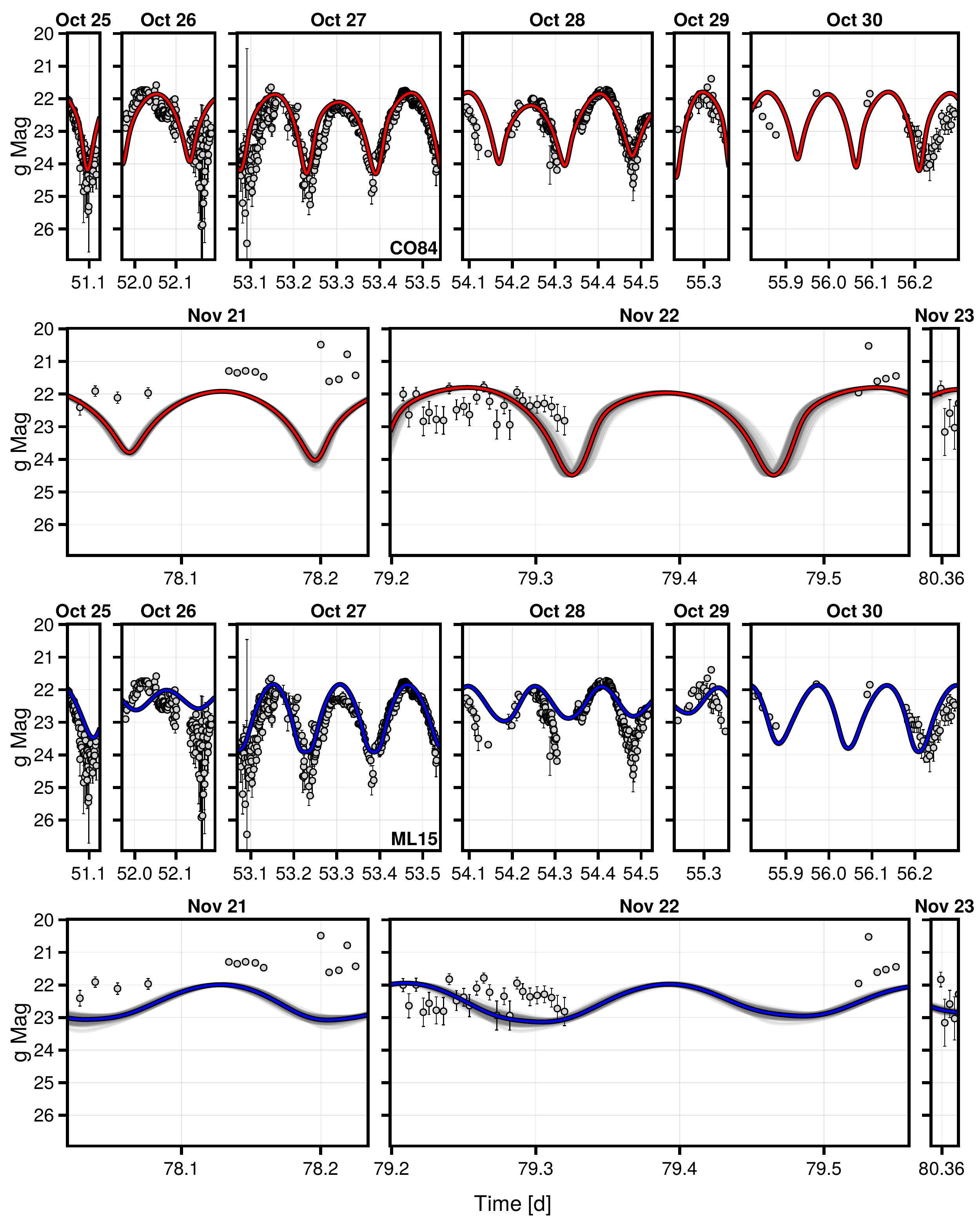}
\caption{An ensemble of torque models, sampled from a multivariate Gaussian distribution centered around the optimal parameter values with means and standard deviations given by Table \ref{table:optimalparams}. They are plotted versus the photometric data (gray points) taken in October 2017 (1st and 3rd rows) and November 2017 (2nd and 4th rows). The curves (in the first two rows) are the optimal fit, and the gray curves are the sampled ensemble. Both the CO84 (red, Equation \eqref{eq:COmodel}) and the ML15 (blue, Equation \eqref{eq:MLmodel}) are plotted, in the first two and last two rows respectively. The x-axis scale for the October 2017 and November 2017 data are different. Note that the ``optimal fit" is not optimal for the November 2017 data. {The time axis has units of days, with a zero point at 58000 MJD.} }
\label{fig:totensemble}
\end{figure*}

In this subsection, we evolve the optimal rotation model found in Section \ref{subsec:oumuamuafit} over `Oumuamua's trajectory until it is at 15 au. We calculate the evolution of the rotation state and the angular momentum over time. This model is run both forwards and backwards in time from the observation dates. We reverse the sign of the time step in our model for backwards modeling. 

It is important to note that H$_2$O outgassing is generally inactive beyond the orbit of Jupiter \citep{Jewitt2009}, {and falls off rapidly beyond 3 au}, well within the boundaries of this extension. However, the H$_2$ outgassing mechanism presented in \citet{Bergner2023} operates inwards of $\sim$30 au (see Figure 3 in that paper). Therefore, the results of this section only apply under the assumption of hypervolatile outgassing (H$_2$, N$_2$, CO) for `Oumuamua (to the inner Centaur region at 5 au, \citet{Gladman2008,Jewitt2009}) or the model of \citet{Bergner2023} further out.

In Figure \ref{fig:angmommag}, we show the average rotational period versus heliocentric distance. The weaker torque in the ML15 optimum causes a smaller variation in the angular momentum {magnitude}. The central spike appears because of the increased force close to the Sun, which causes a temporary spin-up. Notably, the substellar outgassing model has no secular effect on the angular momentum. The angular momentum is generally conserved as `Oumuamua enters and exits the solar system. Similar behavior is produced by ensemble models (not shown), confirming the robustness of the result given the idealized assumptions of the model. {We compare the rotation state of these models to a rotation state about the smallest principal axis. We present the angle between the angular velocity vector and the smallest principal axis, $\theta_{\rm NPA}=\arccos(\boldsymbol{\hat{\omega}}\cdot\boldsymbol{\hat{z}})$, to parameterize this. We find that for the CO84 model, the initial state is $\theta_{\rm NPA,0}=-1.51\pm2.7^\circ$ and the final state is $\theta_{\rm NPA,f}=-0.36\pm5.4^\circ$. For the ML15 model, the initial state is $\theta_{\rm NPA,0}=-9.12\pm2.3^\circ$ and the final state is $\theta_{\rm NPA,f}=14.10\pm1.4^\circ$.}

We also compare these synthetic models to data collected in November 2017. These data are quite sparse, and so they are not included in our optimization for the rotation parameters. We generate 100 values of the parameters from a Gaussian distribution with the uncertainties given in Table \ref{table:optimalparams}. We then plot this ensemble of models versus the photometric data in Figure \ref{fig:totensemble}. Note that the curves in Figure \ref{fig:totensemble} are best-fit to the October 2017 data only, not the November 2017 data. It is clear that, as expected, the models diverge over longer integration times. 
 
\section{Discussion}\label{sec:discussion}

{In this work, we expand upon the work of \citet{Mashchenko2019}, incorporating a time-varying outgassing torque which tracks the substellar point, which we find accurately reproduces the photometric light curve.}

By evolving the rotation state into pre-detection, we find that `Oumuamua's initial (and final) rotation states are nonprincipal axis rotations, with little secular change of angular momentum over `Oumuamua's solar system flyby. The initial and final rotation states of `Oumuamua exhibit a relatively rapid rotation, {close to the shortest principal axis}. This rotation state is unmodified by `Oumuamua's close solar encounter. It is of course possible that the assumption of a substellar outgassing jet is a poor model for this object, which would enable an outgassing spin-up. {It is also worth noting that regardless of volatile species or torque model, outgassing would necessarily change the size, shape, and mass of `Oumuamua over its solar system trajectory. \citet{SL2020} showed that `Oumuamua's aspect ratio can change by up to a factor of 3 and the mass by a factor of 5, which is not incorporated in our model.} 

Due to the sparseness of data for November 2017, we are unable to fit a light curve to this observational period. However, the best-fitting rotational state for the October 2017 data does not match the November 2017 data, as shown in Figure \ref{fig:totensemble}. {Note that for most of UT 2017-11-21, the photometric observations are approximately 1 magnitude greater than the predicted values. This could be due to an error in the scattering functions or a resurfacing in the object due to outgassing, increasing the albedo.} Although the lack of data makes differences difficult to confirm, this result indicates that {some modulation of the rotational state has occurred. This may reflect a change in the torques due to outgassing, the properties of  body itself, or  the incompleteness of the best-fit model}. Note that in a previous version of this manuscript, we performed fitting of the November data, which produced spurious results. More data in November 2017 would allow for constraints on the rotation parameters during that time period. These results motivate high-cadence photometric observations over longer temporal baselines for future interstellar objects. 

It is important to note that inaccuracies remain in the light curve models presented in this paper. For example, certain predicted light curve minima are temporally offset from the photometric minima (note the minimum at 52.25 d in Figure \ref{fig:optmodelfits}). Predicted minima are also generally shallower than those observed (again see Figure \ref{fig:optmodelfits}). This inaccuracy has several possible causes, which must be addressed (although outside of the scope of this work) in order to improve our understanding of small-body outgassing and rotational evolution. Firstly, `Oumuamua's shape is likely not a perfect ellipsoid, which would introduce discrepancies in the torque model and the photometric magnitude variations. {In addition, we use shape parameters from \citet{Mashchenko2019}, despite that author using a different light curve model and torque model. The incorporation of ellipsoid shape parameters into the models that we use would likely improve the accuracy, although the computational expense renders this beyond the scope of this paper.} Tidal torques could also contribute to the rotation state. While we demonstrated in Section \ref{sec:bigana} that order-of-magnitude scaling analysis suggests that outgassing is the dominant contributing factor, we identify pathological cases where both forces are comparable. 

A more likely explanation is inaccuracy in the assumptions used in the outgassing model. While this substellar outgassing jet is a highly simplified and idealized model, there is some empirical motivation for its validity. The comet 67P/Churyumov-Gerasimenko exhibited an outgassing which closely follows the substellar point \citep{Kramer2019}, providing a motivation for this idealized model. While the outgassing behavior of `Oumuamua is unknown, the substellar model is similar to the outgassing of at least one comet for which \textit{in situ} measurements exist \citep{Taylor2017}. Despite the relevance of the Sun-tracking jet to 67P, empirical results for larger samples of comets have demonstrated that outgassing torques are generally responsible for nucleus spin-up \citep{Jewitt2003b,Drahus2011,Gicquel2012,Maquet2012,Fernandez2013,Wilson2017,Eisner2017,Roth2018,Kokotanekova2018,Biver2019,Combi2020,Jewitt2021,Jewitt22} {or spin-down \citep{Eisner2017,Bodewits2018,Farnham2021}}. 

{In addition, our} understanding of outgassing mechanics is incomplete, and there are many mechanisms (such as full-surface outgassing, time delays, and stochastic outgassing such as that observed in 103P/Hartley 2, \citet{AHearn2011}) which were not accounted for in our idealized model. Note that \citet{SLB2019} performed numerical experiments with time delays and stochastic forcing of the idealized Sun-tracking jet and also reported a lack of secular spin-up.  `Oumuamua's absence of a visible coma prevents observations of stochastic outgassing (although the sparseness of the astrometric data prohibits differentiation between stochastic and continuous acceleration, \citet{seligman2021}), creating significant difficulty in fitting the light curve. 

{The recent discovery of ``dark comets'' in the solar system \citep{Chesley2016,Farnocchia2023,Seligman2023} --- {comaless} objects with significant non-radial nongravitational accelerations too strong to be explained by non-outgassing effects --- provides an excellent opportunity to resolve outstanding questions about the nature of undetected outgassing. Future investigations of these comets with JWST and with the {Hayabusa2 extended mission to the ``dark comet" 1998 KY26} \citep{Hirabayashi2021} could measure the predicted levels of outgassing and possibly dust.}

`Oumuamua left many unanswered questions as it exited the Solar System, and there is still no general consensus regarding the provenance of the object. The discovery implies a spatial number density of similar objects of order $n_{o}\sim1-2\times 10^{-1}\,$au$^{-3}$ \citep{Trilling2017,Laughlin2017,jewitt2017,moro2018,Zwart2018,Do2018,moro2019a}. Detection and characterization of future interstellar objects offer the most promising avenue for resolving these questions. The forthcoming Rubin Observatory Legacy Survey of Space and Time (LSST) \citep{jones2009lsst,Ivezic2019} will effectively detect such transient objects \citep{solontoi2011comet,Veres2017,veres2017b,Jones2018}, detecting approximately 1 interstellar object every year \citep{Moro2009,Engelhardt2014,Cook2016,Trilling2017,Seligman2018,Hoover2022}. In addition, the forthcoming NEO Surveyor \citep{Mainzer2015} may also detect interstellar objects, and its infrared capabilities could offer information about outgassing sources. Space-based \textit{in situ} measurements of an interstellar object would also provide valuable information regarding the composition and rotation state \citep{Hein2017,Seligman2018,Meech2019whitepaper,Castillo-Rogez2019,Hibberd2020,Donitz2021,Sanchez2021,Meech2021,Hibberd2022,Moore2021whitepaper,Moore2021}.

\section{{Summary \& Conclusions}}\label{sec:conclusion}

{In this paper, we developed a method to generate synthetic light curves of NPA rotators under the action of outgassing torques.  We introduced a  numerical rotation model for these objects that incorporates torques imposed by substellar outgassing. We then analyzed the rotation state of `Oumuamua using this methodology, with the magnitude of the outgassing acceleration set by the nongravitational acceleration found by \citet{micheli2018}.}

{This work expands upon the work of \citet{Mashchenko2019}, which produced a rigorous fit to the photometric light curve using a similar rotation model. In that work, it was shown that the observed data are difficult to accurately reproduce in the absence of a torque (which we confirm, see Appendix \ref{sec:zerotorquemodel}). Our most significant addition to the work of \citet{Mashchenko2019} is the incorporation of a time-varying outgassing torque, which we assume to be normal to the surface and directed from the substellar point. This torque is based on the reported nongravitational acceleration of `Oumuamua. This model visually reproduces the accuracy of \citet{Mashchenko2019}'s model (see Figure 6 in that manuscript). We also extend this model beyond the observed time frame to investigate the rotation state both pre- and post-observation. Our main findings are as follows: }

{(i) We compute and report parameters for both a geometric scattering model \citep{Connelly1984} and a Lommel-Seeliger scattering model \citep{muinonen2015}, which produce light curves similar to the photometric data. }

{(ii) We find that under the assumptions of our models, `Oumuamua does not exhibit a secular change of angular momentum over its solar system flyby. We find that the change in the rotation period was $0.31$ hr for the CO84 model, and $-0.09$ hr for the ML15 model. The spin-down reported by \citet{flekkoy2019} may therefore be an artifact of the data sampling, which can produce an apparently varying angular momentum magnitude from torque-driven periodic variability. \citet{Mashchenko2019} discussed the issues with interpreting a period from an object with an applied torque, as the forcing can cause physical frequencies to be obscured in periodogram analysis due to the period modulation.}

{(iii) We demonstrate that under the assumption of a substellar point model, `Oumuamua's likely entered the solar system with rapid rotation about the smallest principal axis. The angle between the rotation axis and the smallest axis, $\theta_{\rm NPA}$, is $-1.51\pm2.7^\circ$ for the CO84 model and $-9.12\pm2.3^\circ$ for the ML15 model. This rotation state may result from damping of a dynamical NPA rotation over long time scales \citep{Burns1973,Sharma2005,Breiter2012}. The damping time scale is $\sim$1 Gyr for an object like `Oumuamua \citep{drahus2017}. This rotation state is maintained when `Oumuamua exited the solar system, with $\theta_{\rm NPA}=-0.36\pm5.4^\circ$ for the CO84 model and $\theta_{\rm NPA}=14.10\pm1.4^\circ$ for the ML15 model. Both models, however, show some destabilization, with the average angle or the variation increasing.} 

{(iv) The rapid rotation of `Oumuamua during the perihelion passage has implications for the stability of the object. For a density of $\rho\sim1$ g cm$^{-3}$, the strength-free lower limit on the rotation period ranges from $\sim3.3-8$ hr. The high rotation rate during perihelion of $\sim 4$ hr for CO84 and $\sim 2$ hr for ML15 implies that `Oumuamua is either not a strength-free body or has a relatively high density, if it did not disintegrate during its perihelion passage, as suggested by \citet{Sekanina2019b}.}

{(v) The light curve models presented in this paper have several drawbacks which prevent this work from being conclusive. A greater understanding of outgassing and nongravitational accelerations would be necessary to effectively analyze both `Oumuamua and future interstellar objects. }

\section*{Acknowledgements}
We thank Douglas MacAyeal for essential help at the outset of this study, for discussion at the time revisions were made, and for providing funding through NSF award 1841467 to the University of Chicago. {We thank Sergey Mashchenko for his detailed suggestions regarding this manuscript after our initial submission.} {We thank the two anonymous reviewers for their helpful and insightful comments that strengthened the scientific content of this manuscript.} We thank Adina Feinstein, David Jewitt, and Faith Vilas for useful conversations and suggestions. We thank the two original anonymous reviewers for their constructive feedback on the scientific content of this paper, especially on the suggestion to split the work into two manuscripts. DZS acknowledges financial support from the National Science Foundation Grant No. AST-2107796, NASA Grant No. 80NSSC19K0444 and NASA Contract  NNX17AL71A. {Simulations in this paper made use of the REBOUND N-body code \citep{rebound}. The simulations were integrated using IAS15, a 15th order Gauss-Radau integrator \citep{reboundias15}. }

\appendix

\begin{figure*}
\centering
\includegraphics[width=\linewidth,angle=0]{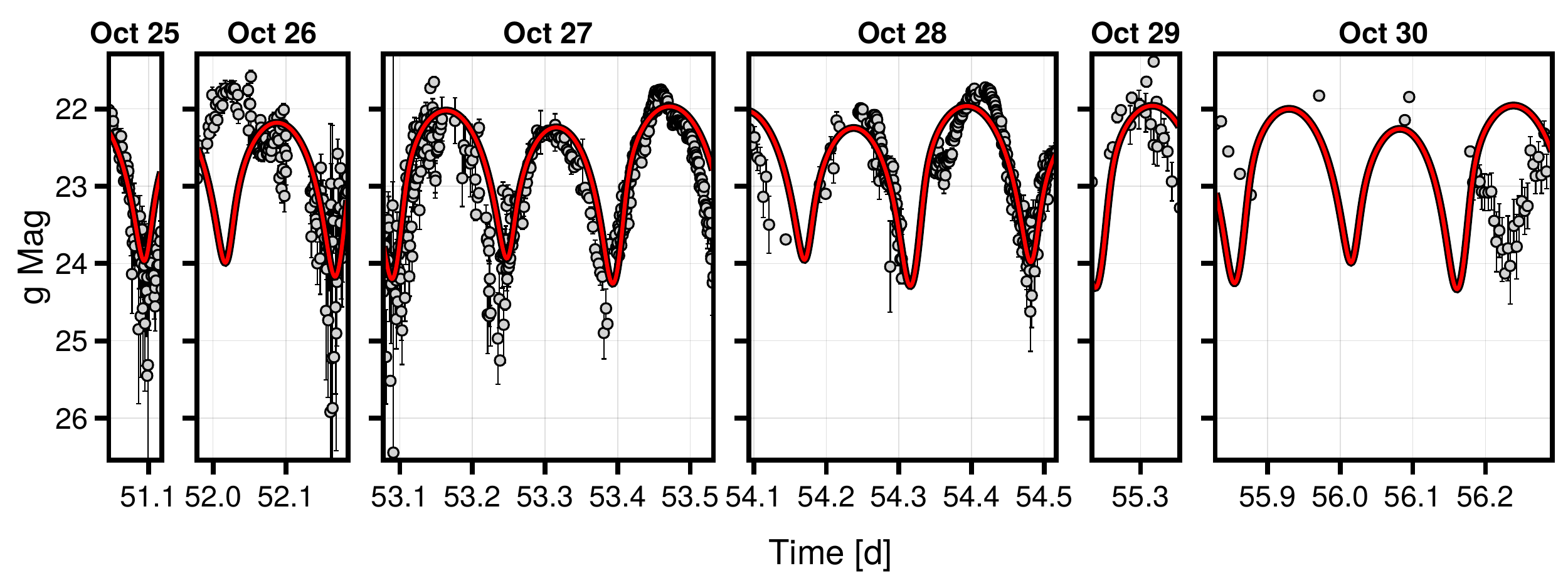}
\caption{ {Best-fit synthetic light curves (red line) and photometric data (grey points) for a torque-free model, using the CO84 light curve model. The data shown are those taken in October 2017 only. The parameters for the model are given in Table \ref{table:torquefreeparams}. {The time is in days, with a zero point at 58000 MJD.}} }
\label{fig:torquefreefig}
\end{figure*}

\section{Dumbbell Model Validation}\label{sec:dumbval}

In this appendix, we validate the `dumbbell' model {used in Section \ref{sec:bigana}} to represent the permanent quadrupole moment of `Oumuamua. We compare the behavior of Equation \eqref{eq:tidal} with numerical integrations of tidal torques over a continuous body. 

The continuum force per unit volume is given by 
\begin{equation}
    \boldsymbol{F}_{\text{tidal}}=- GM_{\Sun}\rho\left(\frac{1}{(r_H-x)^2}-\frac{1}{r_H^2}\right) \,\big(\,\boldsymbol{R}\cdot\boldsymbol{\hat{x}}\,\big)\,.
\end{equation}
The torque is then $\boldsymbol{\tau}=\boldsymbol{d}\times\boldsymbol{F}$, where the distance from the rotation axis $\boldsymbol{d}$ is given by
\begin{equation}
    \boldsymbol{d}=\boldsymbol{r}-(\boldsymbol{r}\cdot\boldsymbol{\hat{\Omega}})\boldsymbol{\hat{\Omega}}\,.
\end{equation}
Here, $\boldsymbol{r}$ is a position vector, and $\boldsymbol{\hat{\Omega}}$ is the rotation axis. 

The torque on `Oumuamua is evaluated with Equation \eqref{eq:tidal} on a $100\times500$ grid for $\beta\in[0,\pi)$ and $r_H\in[0.257,3.147]$ au, with $\boldsymbol{\hat{\Omega}}=(0,1,0)$. Numerical integrals are computed over this grid with the \texttt{quadpy} package \citep{quadpy}. The integral domain is the set of $(u',v',w')$ Cartesian points in the unit 3-ball $\mathcal{S}$, transformed via
\begin{equation}
    \begin{dcases}
        x'=\frac{a}{2}u'\\
        y'=\frac{b}{2}v'\\
        z'=\frac{c}{2}w'\,.
    \end{dcases}
\end{equation}
These relationships define an isomorphic mapping $(x',y',z')\rightarrow(u',v',w')$ from $\mathds{R}^3\rightarrow\mathds{R}^3$. The determinant of the Jacobian is $du'dv'dw'=8abc\,dx'dy'dz'$, which finalizes the transformation. We define the fractional error as $E_{\rm frac}=|1-D/M|$ for numerical ($D$) and analytic ($M$) values.  The fractional error is everywhere less than $10^{-3}$, except at $\beta=\pi/2$, where it reaches unity. 

The value of unity corresponds to where the numerical solution is significantly smaller than the analytical solution. By inspection, the torque here is exactly 0, but floating-point imprecision produces orders-of-magnitude difference. Both methods return torques with magnitudes close to 0 at this point ($\mathcal{O}(10^{-8})$), so it is only their ratio that is large. We also verified that the error has no dependence on heliocentric distance $r_H$. Therefore, the `dumbbell' model accurately depicts the torque on an ellipsoidal body, with errors of $\mathcal{O}(10^{-4})$. 

\section{Cigar-shape Tidal Torque}\label{sec:cigtor}

In this appendix, we demonstrate that the tidal torque is equal for a `cigar'- and `pancake'-shaped body (Section \ref{subsec:tidana}). We assume now that $b=c$, $a>c$. Again equating the quadrupole moments, we find that the points are at $(\pm R,0,0)$, once again with $R\equiv\sqrt{a^2-c^2}$ and the mass $m=(4\pi/15)abc\rho$. We again assume rotation about the y-axis, so when the point masses are rotated about the x-axis by an angle $\beta$, the masses are at $\pm (R\cos\beta,0,-R\sin\beta)$. As before, the tidal acceleration on each of the points is given by Equation \eqref{eq:torqueforce}, and the lever arm is $\pm (R\cos(\beta),0,-R\sin(\beta))$ for $m_1$ and $m_2$ respectively. Therefore, the resulting tidal torque is 
\begin{equation} 
\begin{aligned}
    \boldsymbol{\tau}_{m_1/m_2}=- &G M_{\Sun} m \left( \frac{1}{(r_H\mp R\cos\beta)^2}-\frac{1}{r_H^2} \right) \\ &[(1,0,0)\times\pm(R\cos\beta,0,-R\sin\beta)]
\end{aligned}
\end{equation} 
The vector component is $\mp R\sin\beta\,\boldsymbol{\hat{y}}$. Simplifying using $m=(4\pi/15)abc\rho$, $R=\sqrt{a^2-c^2}$, $b=c$, and $\epsilon=c/a$, the magnitude of the total torque along the rotation axis (which is once again assumed to be $\boldsymbol{\hat{y}}$) is
\begin{equation}
    \tau=(\frac{8\pi}{15}GM_{\Sun})\left(\frac{r_Ha^5\rho\epsilon^2(1-\epsilon^2)\sin{(2\beta)}}{(r_H^2-a^2(1-\epsilon^2)\cos^2(\beta))^2}\right ).
\end{equation}
Note that apart from an $\epsilon^2$ to account for the different sizes of $b$, this is equivalent to Equation \eqref{eq:tidal}, confirming the applicability of this model.

\section{{Torque-Free Geometric Light Curve Model}}\label{sec:zerotorquemodel}

{In this section, we obtain an optimal rotation state for `Oumuamua, using the geometric scattering model (CO84, see Section \ref{subsec:geomag}) in the absence of outgassing-induced torque. A similar computation for the Lommel-Seeliger scatting model was given in \citet{Mashchenko2019}, and we therefore only present results for the CO84 model.  }

\begin{table}[ht]
\centering
\caption{{The optimal initial rotation parameters for `Oumuamua, using the CO84 model (Section \ref{subsec:geomag}) in the absence of outgassing-induced torque. The $1\sigma$ uncertainty expressed for the parameters is given by the Cramér-Rao bound, which is obtained by fitting a Gaussian likelihood model to the region around the maximum. The given frequency $\nu$ is the instantaneous rotation frequency at the time of first observation, which corresponds to the reciprocal of the period. To enable comparison to other models, we also report the $\chi^2$ values for each model fit. Note that the $\chi^2$ value and the log-likelihood are co-optimal. }}
\begin{tabular}{ cr } 
 {Parameter:} & \multicolumn{1}{c}{{CO84:}} \\ 
 \hline
 {$\alpha$ [rad]} & {$\phantom{-}0.8394\pm4.0\times10^{-3}$} \\
 {$\beta$ [rad]} & {$\phantom{-}3.0792\pm3.4\times10^{-3}$} \\
 {$\gamma$ [rad]} & {$-0.7692\pm2.2\times10^{-3}$} \\
 {$\theta$ [rad]} & {$\phantom{-}2.1922\pm1.1\times10^{-4}$} \\
 {$\phi$ [rad]} & {$\phantom{-}2.5634\pm6.5\times10^{-5}$} \\
 {$\nu$ [hr$^{-1}$]} & {$\phantom{-}0.1085\pm3.7\times10^{-6}$} \\\hline
 {$L$ [$J_a$ d$^{-1}$]} & {$21.1653\pm9.9\times10^{-4}$} \\
 {$\theta_L$ [rad]} & {$0.1101\pm2.8\times10^{-3}$} \\
 {$\varphi_L$ [rad]} & {$5.1259\pm2.0\times10^{-2}$} \\
 {$\varphi_0$ [rad]} & {$0.5640\pm2.9\times10^{-3}$} \\
 {$\psi_0$ [rad]} & {$6.1566\pm2.3\times10^{-3}$} \\
 {$E'$ [$J_a^{-1}$]} & {$0.4311\pm2.1\times10^{-5}$} \\\hline
 {$\chi^2$} & \multicolumn{1}{c}{{31.5935}} \\
\end{tabular}
\label{table:torquefreeparams}
\end{table}

{In this section, we follow the same procedure as in Section \ref{subsec:oumuamuafit}, although we forcibly set the outgassing jet scaling parameter $f=0$. The best-fit parameters obtain are presented in Table \ref{table:torquefreeparams}, and the optimal synthetic light curve is shown in Figure \ref{fig:torquefreefig}. For this rotation state, the $\chi^2$ value is 31.5935, significantly larger than the value of 22.0418 (Table \ref{table:optimalparams}) for the model including torque. In addition, the light curve is visually inferior to those in Figure \ref{fig:optmodelfits}, with an additional sharp minimum at a time of 52 d, and a poor fit to the data from October 30 2017. As a result, we conclude that torque is necessary to explain the rotation state of `Oumuamua. }

\bibliography{main}{}
\bibliographystyle{aasjournal}

\end{document}